\documentclass[12pt]{article}

\usepackage[dvips]{graphicx}
\usepackage[dvips]{color}

\advance\textwidth by 100pt
\advance\oddsidemargin by -50pt
\advance\evensidemargin by -50pt
\newcommand{\eq}[1]{(\ref{#1})}
\def \hc {{\rm h.c.}}

\def\wt{\widetilde}
\renewcommand{\thesection}{\arabic{section}}
\renewcommand{\theequation}{\thesection.\arabic{equation}}
\newcommand{\cleqn}{\setcounter{equation}{0}}

\newcommand {\hate}     {\hat{e}}
\newcommand {\hatg}     {\hat{g}}
\newcommand {\hats}     {\hat{s}}
\newcommand {\hatc}     {\hat{c}}
\newcommand {\hatgz}    {\hat{g}_Z}

\newcommand {\hatesq}   {\hat{e}^2}
\newcommand {\hatgsq}   {\hat{g}^2}
\newcommand {\hatssq}   {\hat{s}^2}
\newcommand {\hatcsq}   {\hat{c}^2}

\newcommand {\msbar} {\overline{\rm MS}}
\newcommand {\mwsq}  {m_W^2}

\newcommand {\pitww}    {\Pi_T^{WW}}

\newcommand {\pitggg} {{\Pi}_{T,\gamma}^{\gamma\gamma}}

\newcommand {\pitgzg} {{\Pi}_{T,\gamma}^{\gamma Z}}
\newcommand {\pitzzz} {{\Pi}_{T,Z}^{ZZ}}

\newcommand {\qsq} {{q^2}}

\newcommand {\boxes}  {{\rm [Box]}}

\def\mn#1{m_{\widetilde{\chi}^0_#1}}
\def\mch#1{m_{\widetilde{\chi}^-_#1}}
\def\cunitary#1{U^C_#1}
\def\nunitary#1{U^N_#1}
\def\half{\frac{1}{2}}

\def\ehat{\hat{e}}
\def\shat{\hat{s}}
\def\gzhat{\hat{g}_Z^{}}
\def\gzhatsq{\hat{g}_Z^2}

\def\to{\rightarrow}

\def\ov{\overline}

\def\eeww{e^+ e^- \to W^+ W^-}
\def\eewo{e^+ e^- \to \omega^+ W^-}

\def\mw{m_W^{}}

\newcommand{\beq}{\begin{equation}}
\newcommand{\eeq}{\end{equation}}
\newcommand{\bea}{\begin{eqnarray}}
\newcommand{\eea}{\end{eqnarray}}
\newcommand{\bsub}{\begin{subequations}}
\newcommand{\esub}{\end{subequations}}
\renewcommand{\theequation}{\thesection.\arabic{equation}}

\def\PRD#1#2#3{Phys. Rev. {\bf D#1} (19#2) #3}
\def\PRDT#1#2#3{Phys. Rev. {\bf D#1} (20#2) #3}
\def\NPB#1#2#3{Nucl. Phys. {\bf B#1} (19#2) #3}

\def\ZPC#1#2#3{Z. Phys. {\bf C#1} (19#2) #3}
\def\EPJC#1#2#3{Eur. Phys. J. {\bf C#1} (19#2) #3}
\def\EPJCT#1#2#3{Eur. Phys. J. {\bf C#1} (20#2) #3}
\def\PLB#1#2#3{Phys. Lett. {\bf B#1} (19#2) #3}

\makeatletter
\newtoks\@stequation

\def\subequations{\refstepcounter{equation}%
  \edef\@savedequation{\the\c@equation}%
  \@stequation=\expandafter{\theequation}
  \edef\@savedtheequation{\the\@stequation}
  \edef\oldtheequation{\theequation}%
  \setcounter{equation}{0}%
  \def\theequation{\oldtheequation\alph{equation}}}

\def\endsubequations{%
  \ifnum\c@equation < 2 \@warning{Only \the\c@equation\space subequation
    used in equation \@savedequation}\fi
  \setcounter{equation}{\@savedequation}%
  \@stequation=\expandafter{\@savedtheequation}%
  \edef\theequation{\the\@stequation}%
  \global\@ignoretrue}

\def\eqnarray{\stepcounter{equation}\let\@currentlabel\theequation
\global\@eqnswtrue\m@th
\global\@eqcnt\z@\tabskip\@centering\let\\\@eqncr
$$\halign to\displaywidth\bgroup\@eqnsel\hskip\@centering
     $\displaystyle\tabskip\z@{##}$&\global\@eqcnt\@ne
      \hfil$\;{##}\;$\hfil
     &\global\@eqcnt\tw@ $\displaystyle\tabskip\z@{##}$\hfil
   \tabskip\@centering&\llap{##}\tabskip\z@\cr}

\makeatother

\begin{document}
\thispagestyle{empty}
\vspace*{-15mm}
\baselineskip 10pt
\begin{flushright}
\begin{tabular}{l}
{\bf KEK-TH-857}\\
{\bf DESY 02-205}\\
{\bf PITHA 02/16}\\
{\bf hep-ph/0212135}
\end{tabular}
\end{flushright}
\baselineskip 24pt 
\vglue 15mm 
\begin{center}
{\Large\bf One-loop contributions of charginos and neutralinos to\\
         $W$-pair production in $e^+e^-$ collisions \\ 
}
\vspace{5mm}

\baselineskip 18pt 
\def\thefootnote{\fnsymbol{footnote}}
\setcounter{footnote}{0}

{\bf 
Kaoru Hagiwara$^{1}$,
Shinya Kanemura$^1$, 
Michael Klasen$^{2}$\footnote{Supported by DFG under contract Nr.
KL 1266/1-3.} 
and 
Yoshiaki Umeda$^{3}$\footnote{Supported by BMBF, contract nr. 05
HT1PAA 4.}}

\vspace{5mm}
{\it 
$^1$Theory Group, KEK, Tsukuba, Ibaraki 305-0801, Japan\\
$^2$II Institut f\"{u}r Theoretische Physik, Universit\"{a}t Hamburg,
 D-22761 Hamburg, Germany\\
$^3$Institut f\"{u}r Theoretische Physik E, RWTH Aachen,
 52056 Aachen, Germany}
\end{center}


\vspace*{15mm}

\begin{center}
\begin{abstract}
\vspace*{4mm}
\begin{minipage}{15cm}
\rm

We study the one-loop effects of charginos and neutralinos on the 
helicity amplitudes for $\eeww$ in the minimal supersymmetric 
standard model.  
The calculation is tested by using two methods.  
First, the sum rule for the form factors 
between $\eeww$ and the process where the external $W^\pm$ bosons 
are replaced by the corresponding Goldstone bosons $\omega^\pm$ 
is employed to test the analytic expression and  
the accuracy of the numerical program. 
Second, the decoupling property in the large mass limit is used 
to test the overall normalization of the amplitudes. 
These two tests are most effectively carried out 
when the amplitudes are expanded in terms of the 
modified minimal subtraction 
($\overline{\rm MS}$ )
couplings of the standard model.
The resulting perturbation expansion is valid at collider energies 
below and around the threshold of the light supersymmetric particles.
We find that the corrections to the cross section of the longitudinally 
polarized $W$-pair production can be as large as $-1.4$\% at the 
threshold of the light chargino-pair production for large scattering angles. 
We also study the effects of the $CP$-violating phase in the chargino 
and neutralino sectors on the helicity amplitudes. 
We find that the resulting $CP$-violating asymmetries can be at most 
0.1\%.

\end{minipage}
\end{abstract}
\end{center}

\newpage
\baselineskip 16pt 
\def\thefootnote{\arabic{footnote}}
\setcounter{footnote}{0}
\section{Introduction}
\cleqn

\hspace*{12pt}

$W$-boson-pair production has been the benchmark process of  
the CERN $e^+$$e^-$ collider
LEP2, and will continue being so at future linear $e^+e^-$ collider 
experiments because of its large production rate and its possible 
sensitivity to the physics of electroweak symmetry breakdown.
At linear colliders, precise measurements of the masses of the 
$W$ boson, top quark,
and possibly the Higgs boson will be achieved, and there is  
hope of detecting new physics signals through radiative corrections in
the triple gauge boson ($WW\gamma$ and $WWZ$) vertices. 
In particular, if nature is described by the model with weak scale 
supersymmetry (SUSY), radiative corrections due to 
supersymmetric particles are expected. 

In this paper, we show the one-loop effects of charginos and 
neutralinos on the helicity amplitudes of on-shell $W$-pair production 
in the minimal supersymmetric standard model (MSSM).  
A study of the contribution from squarks and sleptons 
has been reported in Ref.~\cite{eeww_sf}, 
and the range of one-loop corrections in the MSSM has been 
studied in the literature~\cite{hahn}.

In Sec.~2, we review the essential aspects of the form-factor 
formalism and the helicity amplitudes for the process $\eeww$. 
A form-factor decomposition of helicity 
amplitudes~\cite{gg79,hpzh87,jeger} 
is useful to calculate the one-loop 
effects, and hence we present our result by extending the 
formalism of Ref.~\cite{hhis96} such that the unphysical scalar 
polarization of the final-state $W$ bosons can also be 
studied~\cite{hisz93,brs}.  
These scalar polarization contributions and the process including
the Nambu-Goldstone boson ($e^+e^- \rightarrow \omega^+\,W^-$) 
are necessary to perform the test by using the 
Becchi-Rouet-Stora (BRS) sum rules~\cite{brs}. 

In Sec.~3, the one-loop chargino and neutralino effects on the gauge 
couplings, the weak boson masses, and the form factors are presented 
in the modified minimal subtraction
($\overline{\rm MS}$) scheme~\cite{msbar}.
In Sec.~4, our one-loop calculation for the amplitude is tested by using 
the BRS sum rule and the decoupling property. 
First, the BRS sum rule for the form factors between $\eeww$ and 
$e^+e^- \rightarrow \omega^+\,W^-$ 
is used to test the analytic expressions and the accuracy of 
the numerical program. 
This test is useful in the process $\eeww$ where 
the gauge theory cancellation among one-loop diagrams 
becomes severe at high energies. 
We confirm numerically that the form factors satisfy the BRS sum 
rule within the expected accuracy of the numerical program.
Second, the decoupling property in the large mass limit is used 
to test the normalization of the amplitudes. 
By expanding the one-loop amplitudes in terms of the 
$\msbar$ couplings of the SM, the decoupling of the SUSY 
particle effects is made manifest in the large mass limit. 
This test ensures the validity of the renormalization scheme
and confirms the overall normalization factors  
such as the external wave-function contribution, which cannot be
tested by the BRS sum rules. 
We find that the above two tests are most effectively carried out 
when the amplitudes are expanded in terms of the $\overline{\rm MS}$ 
couplings of the standard model.
The resulting perturbation expansion is valid at collider energies 
below and around the light SUSY particle thresholds.

In Sec.~5, we present a numerical 
study of the $\eeww$ helicity amplitudes. 
We also examine the effects of the $CP$-violating phases of the chargino 
and neutralino sector. In Sec.~6 we present our conclusion.

In Appendix  A, we summarize our notation for the mass terms 
and the interactions of the chargino and neutralino sector of the MSSM.
The formulas for the one-loop contributions to 
the two-point functions and the vertex functions 
are listed in Appendix B.

\section{The helicity amplitudes}
\cleqn

\subsection{$\eeww$}
\label{sec-eeww-amp}

\hspace*{12pt}
We consider the process
\begin{equation}
e^-(k,\tau) + e^+(\overline{k},\overline{\tau}) \rightarrow
W^-(p,\lambda) + W^+(\overline{p},\overline{\lambda}),  
\end{equation}
where the incoming momenta of $e^-$ and $e^+$ are $k$ and $\overline{k}$ 
as well as the outgoing momenta of $W^-$ and $W^+$ are $p$ and $\overline{p}$, 
respectively.  
The helicity of the incoming $e^-$ ($e^+$) is given by $\frac{1}{2}\tau$ 
($\frac{1}{2}\overline{\tau}$), and 
that of the outgoing $W^-$ ($W^+$)  is given by  
$\lambda$ ($\overline{\lambda}$).
In the limit of massless electrons, only the $\overline{\tau} = -\tau$ 
helicity amplitudes survive. They are written 
for each set of $\{\tau, \lambda, \ov{\lambda}\}$ as~\cite{hhis96,brs} 
\begin{equation}
\label{amp-eeww}
{\cal M}^{\lambda \ov{\lambda}}_\tau
  (\eeww)
= \sum_{i=1}^{16} F_{i,\tau}(s,t)\, j_\mu(k,\overline{k},\tau) 
T_i^{\mu\alpha\beta} \epsilon_\alpha(p,\lambda)^\ast 
\epsilon_\beta(\overline{p},\overline{\lambda})^\ast  \;,
\end{equation}
where all dynamical information is contained in the form factors 
$F_{i,\tau}(s,t)$ with $s = (k + \overline{k})^2 \equiv q^2$ 
and $t = (k - p)^2$.  
The other factors in Eq.~(\ref{amp-eeww}) are of a purely kinematical 
nature; 
$\epsilon_\alpha(p,\lambda)^\ast$ and 
$\epsilon_\beta(\overline{p},\overline{\lambda})^\ast$ are the polarization 
vectors for $W^-$ and $W^+$, respectively, 
and 
$j_\mu(k,\overline{k},\tau)$ is the massless-electron current. 
The 16 independent basis tensors $T^{\mu\alpha\beta}_i$ 
are defined by Eqs.~(2.6) in Ref.~\cite{brs}. 
Processes with physically polarized $W$ bosons 
($\lambda, \ov{\lambda} = -, +$ or $0$) are described 
by the first nine form factors ($i = 1$ to $9$ for $\tau=\pm 1$). 

The 18 physical helicity amplitudes are given in terms of the 
form factors $F_{1,\tau}$ to $F_{9,\tau}$ by 
\footnote{In Ref.~\cite{eeww_sf}, there is a typo in the expression 
for $M^{0\pm}$. The corrected one is given in Eq.~(\ref{m0+}) in this paper.}
\begin{subequations}
\begin{eqnarray}
\!\!\!\!\!\!\!\!\!\!\!\!
M^{00}_{\tau} &=& -s \left[ -\gamma^2\beta (1+\beta^2) F_{1,\tau} 
+ 4\beta^3\gamma^4 F_{2,\tau} +2\beta\gamma^2 F_{3,\tau}
 - 2 \gamma^2 \cos\theta F_{8,\tau}
\right] \sin \theta , \label{m00}\\ 
\!\!\!\!\!\!\!\!\!\!\!\!
M^{\pm 0}_{\tau} &=& s \gamma \left[ \beta (
F_{3,\tau} 
- i  F_{4,\tau} 
\pm        \beta F_{5,\tau}) 
\pm i  F_{6,\tau}  
\pm (\tau \mp 2\cos\theta) F_{8,\tau} 
\mp 4  \gamma^2 \beta \cos \theta F_{9,\tau}
\right] \frac{(\tau \pm \cos\theta)}{\sqrt{2}} , \label{m+0}\\ 
\!\!\!\!\!\!\!\!\!\!\!\!
M^{0\pm}_{\tau} &=& s\gamma\left[ \beta (
F_{3,\tau} + i F_{4,\tau} \mp \beta F_{5,\tau} ) \pm i F_{6,\tau}
\mp (\tau \pm 2\cos\theta)
F_{8,\tau}
\pm 4  \gamma^2 \beta \cos \theta F_{9,\tau}
\right]\frac{(\tau \mp \cos\theta)}{\sqrt{2}} , \label{m0+}\\ 
\!\!\!\!\!\!\!\!\!\!\!\!
M^{\pm\pm}_{\tau}  &=& s\left[
- \beta F_{1,\tau} 
\mp i F_{6,\tau} 
\mp 4 i \beta^2 \gamma^2 F_{7,\tau} 
+ \cos\theta F_{8,\tau}
+ 4 \beta \gamma^2 \tau F_{9,\tau}
 \right] \sin\theta ,\label{m++} \\
\!\!\!\!\!\!\!\!\!\!\!\!
M^{\pm\mp}_{\tau} &=& 
 s ( \mp F_{8,\tau} - 4 \beta \gamma^2 F_{9,\tau}) 
                      (\tau \pm \cos\theta ) \sin\theta ,\label{m+-}
\end{eqnarray}
\end{subequations}
where the scattering angle $\theta$ is measured between the momentum 
vectors of the $e^-$ and $W^-$,  
\begin{equation}
\beta = \sqrt{1-m_W^2/E_W^2}\;,\makebox[.5cm]{} \gamma = E_W/m^{}_W
\;,\makebox[.5cm]{}E_W = \sqrt{s}/2  \;,
\end{equation}
in the center-of-mass frame of $e^+e^-$ collision. 
The properties of $F_{i,\tau}(s,t)$ under the discrete transformations of 
the charge conjugation ($C$), the parity inversion ($P$), and the combined 
transformation $CP$ are summarized in Table~\ref{table-c-p-cp}.  
There are six $CP$-violating form factors 
($F_{4,\tau}^{}$, $F_{6,\tau}^{}$, and $F_{7,\tau}^{}$).

\begin{table}[t]
\begin{center}
\begin{tabular}{|c||c|c|c|c|c|c|c|c|c|}\hline
&\hspace*{0.3cm}$F_1$\hspace*{0.3cm} &\hspace*{0.3cm}$F_2$\hspace*{0.3cm} 
&\hspace*{0.3cm}$F_3$\hspace*{0.3cm} &\hspace*{0.3cm}$F_4$\hspace*{0.3cm} 
&\hspace*{0.3cm}$F_5$\hspace*{0.3cm} &\hspace*{0.3cm}$F_6$\hspace*{0.3cm} 
&\hspace*{0.3cm}$F_7$\hspace*{0.3cm} &\hspace*{0.3cm}$F_8$\hspace*{0.3cm} 
&\hspace*{0.3cm}$F_9$\hspace*{0.3cm} \\ 
\hline \hline
$C$  & $+$ & $+$ & $+$ & $-$ & $-$ & $+$ & $+$ & $+$ & $-$ \\ \hline
$P$  & $+$ & $+$ & $+$ & $+$ & $-$ & $-$ & $-$ & $+$ & $-$ \\ \hline
$CP$ & $+$ & $+$ & $+$ & $-$ & $+$ & $-$ & $-$ & $+$ & $+$ \\ \hline
\end{tabular}
\end{center}
\caption{The properties of the form factors $F_{i,\tau}(s,t)$ under the 
discrete transformations $C$, $P$, and $CP$. Only those that contribute to 
physical processes are listed.}
\label{table-c-p-cp}
\end{table}
The remaining 14 form factors ($i=10$ to $16$ for $\tau=\pm 1$) 
contribute to the amplitudes including unphysical polarizations 
of the $W$ bosons ($\lambda, \ov{\lambda} = S$), 
where the polarization vectors are 
$\epsilon^\alpha(p,\lambda=S)^\ast=p^\alpha/m_W^{}$ and 
$\epsilon^\beta(\overline{p},\overline{\lambda}=S)^\ast
=\overline{p}^\beta/m_W^{}$.

\subsection{$e^+e^- \rightarrow  \omega^+W^-$}

\hspace*{12pt}
To test the $\eeww$ form factors by using the BRS sum rules, 
we also calculate the unphysical process  
\begin{equation}
e^-(k,\tau) + e^+(\overline{k},\overline{\tau}) \rightarrow
W^-(p,\lambda) + \omega^+(\overline{p}),  
\end{equation}
where $\omega^+$ is the Nambu-Goldstone boson associated with 
$W^+$. Our phase convention for $\omega^+$ 
is that of Ref.~\cite{hisz93}.   
We decompose the helicity amplitudes as 
\begin{equation}
 {\cal M}^\lambda_\tau (e^+e^-\rightarrow \omega^+ W^-)
 =    i \sum_{i=1}^{4} H_{i,\tau}(s,t)\, j_\mu(k,\overline{k},\tau) 
S_i^{\mu\alpha} \epsilon_\alpha(p,\lambda)^\ast \;.\label{amp-eewx}
\end{equation}
In Eq. \eq{amp-eewx}, there are four independent basis tensors, 
$S_i^{\mu\alpha}$ ($i=1 - 4$), corresponding to the four (three physical 
plus one scalar) polarizations of the $W^-$ boson.
The form factors are given by $H_{i,\tau}(s,t)$.
The basis tensors $S_i^{\mu\alpha}$ are given in 
Eq. (2.9) of Ref~\cite{brs}.


\section{One-loop chargino and neutralino contributions}
\cleqn

\hspace*{12pt}
In this section, we calculate the one-loop contributions of charginos
and neutralinos to the form factors $F_{i,\tau}^{}$ for $\eeww$ and 
$H_{i,\tau}^{}$ for $e^+e^- \to \omega^+W^-$. 
The Lagrangian for the chargino and neutralino 
sector of the MSSM~\cite{ch99} is given in Appendix A, 
in order to fix our notation. 

\subsection{The renormalization scheme}

\hspace*{12pt}
We explain our renormalization scheme of the MSSM 
parameters, which is designed to make the BRS sum rules 
exact in the one-loop order. 
First, we take the physical $W$ boson mass $\mw$ as one of our input 
parameters as in Ref.~\cite{eeww_sf}.  
The $\overline{\rm MS}$ coupling constants $\hat{e}^2(\mu_R^{})$ and
$\hat{g}^2(\mu_R^{})=\hat{e}^2(\mu_R^{})/\hat{s}^2(\mu_R^{})$ 
of the MSSM are used as the expansion parameters for 
perturbation calculation.
They are obtained from the $\overline{\rm MS}$ couplings of the SM 
by using the matching conditions
\begin{subequations}
\label{eq-mssm}
\begin{eqnarray}
\frac{16\pi^2}{\hat{e}^2 (\mu_R^{})} &=& 
\frac{16\pi^2}{\hat{e}^2_{\rm SM} (\mu_R^{})} - \left[
 \frac{4}{3}\log{\frac{\mu^2_R}{m^2_{\wt{\chi}^{-}_1}}} 
+\frac{4}{3}\log{\frac{\mu^2_R}{m^2_{\wt{\chi}^{-}_2}}}\right],
\label{eq-emssm}\\
\!\!\!\!
\frac{16\pi^2}{\hat{g}^2 (\mu_R^{})} &=& 
\frac{16\pi^2}{\hat{g}^2_{\rm SM} (\mu_R^{})} - 
\left[
 \frac{2}{3}\left\{ (D_L)_{11}+(D_R)_{11} \right\}
           \log{\frac{\mu^2_R}{m^2_{\wt{\chi}^{-}_1}}} 
+\left\{ (D_L)_{22}+(D_R)_{22} \right\}
           \log{\frac{\mu^2_R}{m^2_{\wt{\chi}^{-}_2}}} 
\right]\,,  
\label{eq-gmssm} 
\end{eqnarray}
\end{subequations}
where all the additional particles in the MSSM (squarks, sleptons, 
and extra Higgs bosons) are assumed to be heavy. 
Only the chargino mass  
$m_{\wt{\chi}^{-}_i}$ ($i=1$ and $2$) appears in the matching conditions,  
and the matrices $(D_{\alpha})_{ij}$ that 
relate the weak eigenstates to the mass eigenstates 
are defined in Appendix B~3. 
The numerical results of this report are obtained for 
\begin{eqnarray}
m_W=80.41\; {\rm GeV}, \;\;
\hat{e}_{\rm SM}^{2}(m_Z^{})/(4\pi)=1/128.06,  \;\;
{\rm and} \;\;\hat{s}_{\rm SM}^{2}(m_Z^{})=0.2313, 
\label{eq-w}
\end{eqnarray}  
where the values of $\hat{e}_{\rm SM}^{2}(m_Z^{})$ 
and $\hat{s}_{\rm SM}^{2}(m_Z^{})$ are obtained for $m_t=175$ GeV.
The remaining 
$\overline{\rm MS}$ coupling constants of the SM are then 
calculated in the leading order 
by using Eqs.~(3.5a) and (3.5b) in Ref.~\cite{eeww_sf}.
The above conditions ensure that physical observables at low energies 
remain the same when all the chargino and neutralino masses are large. 
In this paper, we do not consider contributions of sfermions, 
gluinos, or additional Higgs scalar bosons. 
These particles are assumed to be very heavy, and we work within 
the effective MSSM with light charginos and neutralinos. 
The three input parameters $\{ \mw, \hatesq(\mu_R^{}), \hatssq(\mu_R^{}) \}$
are consistently employed in the evaluation of all loop integrals 
and form factors, as well as the chargino and neutralino mixing 
matrix elements.  
All the terms of the relevant diagrams are expanded in powers of 
the $\overline{\rm MS}$ coupling $\hatgsq$ (or $\hatesq$), 
and the terms up to ${\cal O}(\hatg^4)$ are taken into account.

The ${\overline{\rm MS}}$ masses of the weak bosons are calculated 
in the one-loop level as  
\begin{eqnarray}
  \hat{m}_W^2 &=& m_W^2 + \Pi_T^{WW}(m_W^2)      \, , \\
  \hat{m}_Z^2 &=& \frac{\hat{m}_W^2}{\hat{c}^2} 
   = \frac{1}{\hat{c}^2} 
     \left\{ m_W^2 + \Pi_T^{WW} (m_W^2) \right\} \, ,
\end{eqnarray}
where $\Pi_T^{WW}(q^2)$ is the $W$ boson 
two-point function in the $\msbar$ scheme~\cite{hhkm94}, whose 
chargino and neutralino contribution is given in Appendix B~3. 
The $Z$ boson mass is then obtained as 
\begin{eqnarray}
  m_Z^2 = \frac{m_W^2}{\hat{c}^2} 
   + \frac{1}{\hat{c}^2} \Pi_T^{WW} (m_W^2) - \Pi_T^{ZZ} 
 \left(\frac{m_W^2}{\hat{c}^2}\right) \equiv 
  \frac{m_W^2}{\hat{c}^2} + \Delta\; . \label{defmz}
\end{eqnarray}
The chargino and neutralino contributions to the two-point functions 
$\Pi_T^{WW}$ and  $\Pi_T^{ZZ}$ are given in Appendix B~3, and  
the deviation from the tree-level expression is 
denoted by $\Delta$. 
In order to preserve the BRS invariance of the one-loop 
amplitudes exact, the $Z$-boson propagator should be 
expanded and truncated as~\cite{brs} 
\begin{eqnarray}
  \frac{1}{s - m_Z^2} = \frac{1}{s - (m_W^2/\hat{c}^2)} 
     \left\{ 1 + \frac{\Delta}{s - m_W^2/\hat{c}^2}\right\}\;. \label{expmz}
\end{eqnarray}
 
\subsection{One-loop form factors}
\label{subsec-ff-eeww}

\hspace*{12pt}
At the one-loop level, the form factors $F_{i,\tau}(s,t)$, which have been  
introduced in Eq.~(\ref{amp-eeww}), may be written as 
\begin{eqnarray}
  F_{i,\tau} = F_{i,\tau}^{(0)} + F_{i,\tau}^{(1)},  \label{ff01}
\end{eqnarray}
where $F_{i,\tau}^{\rm (0)}$ and $F_{i,\tau}^{\rm (1)}$ are the 
${\cal O}(\hatgsq)$ and ${\cal O}(\hatg^4)$ contributions, respectively. 
We are interested in the $\eeww$ amplitudes for physically 
polarized $W$ bosons ($\lambda, \overline{\lambda} = 0, \pm$). 
In order to test the form factors by using the BRS sum rules,   
we also have to consider the cases in which one or two external $W$ bosons 
have scalar polarization; 
{\it i.e.}, $\lambda$ and/or $\overline{\lambda} = S$.  
Since the BRS sum rules can test the form factors except for the 
overall factors such as the wave-function renormalization 
contribution, we find it convenient to divide the one-loop contribution 
$F_{i,\tau}^{(1)}$ into the following two parts: 
One is the contributions of the external $W$-boson wave-function 
renormalization ($F_{i,\tau}^{\rm (1) ext}$),
and the other is the rest ($F_{i,\tau}^{\rm (1) int}$). 
Equation~(\ref{ff01}) is then rewritten as   
\begin{eqnarray}
  F_{i,\tau} = F_{i,\tau}^{\rm (0)} + F_{i,\tau}^{\rm (1) int} 
             + F_{i,\tau}^{\rm (1) ext} 
             \equiv \widetilde{F}_{i,\tau} + F_{i,\tau}^{\rm (1) ext} .  
    \label{ftilde}
\end{eqnarray}
The explicit forms of $\widetilde{F}_{i,\tau}$ and 
$F_{i,\tau}^{\rm (1)ext}$ are given in Appendix B~1. 
Here, $\widetilde{F}_{i,\tau}$ includes all the one-loop as well as 
tree-level contributions except for the external 
$W$-boson wave-function corrections. This part of the form factors,  
$\widetilde{F}_{i,\tau}$, will be tested by the BRS sum rules in Sec.~4.1, 
while the overall normalization is verified by using the decoupling 
property of the chargino and neutralino contributions in the large 
chargino and neutralino mass limit in Sec.~5.2.  For the BRS test we have to 
calculate all 32 form factors 
$\widetilde{F}_{i,\tau}$ ($i = 1 - 16$) for each $\tau$,  
while we have to calculate the $F_{i,\tau}^{\rm (1) ext}$ only for the 
physical external $W$ lines ($i = 1 - 9$).  
The one-loop level form factors $H_{i,\tau}$ for the process $\eewo$ are 
given in Appendix B~2.

\section{Test of the loop calculation}
\cleqn

\hspace*{12pt}
The purpose of this paper is to evaluate quantitatively 
the one-loop contributions of charginos and neutralinos to the process 
$\eeww$. 
In order to ensure the correctness of our calculation, we 
examine in this section the BRS invariance of our one-loop 
amplitudes and the decoupling behavior of the SUSY effects 
in the large mass limit of charginos and neutralinos.

\subsection{The BRS sum rules
 for the $e^+e^- \rightarrow W^+W^-$ form factors}

\hspace*{12pt}
The standard electroweak theory after gauge fixing is invariant under 
global BRS symmetry, 
so that the amplitudes, that include external massive 
gauge bosons are related to the amplitudes where some of those 
gauge bosons are replaced by their Nambu-Goldstone-boson counterparts.  
From the BRS invariance, the following relations between $\eeww$ 
and $e^+ e^- \rightarrow \omega^+ W^- $ amplitudes
are obtained~\cite{brs,eeww_sf}
\begin{equation}
\label{eq-brs1}
{\cal M}\Big(e^+ e^- \rightarrow  W^+_S\, W^-_P \Big) + i C_{\rm mod}^{BRS}
{\cal M}\Big(e^+ e^- \rightarrow  \omega^+\,W^-_P \Big) = 0\;, 
\end{equation}
where $W_P$ denotes the physical $W$-boson states ($\lambda=\pm 1, 0$)
and $W_S$ denotes its scalar polarization state ($\lambda=S$). 
At loop levels, the factor $C_{\rm mod}^{BRS}$ is not unity, 
and it is found to be~\cite{brs} 
\begin{eqnarray}
C_{\rm mod}^{BRS} = \frac{\hat{m}_W^{}}{m_W^{}}.   \label{cmod-brs}
\end{eqnarray}

By inserting the expressions (2.2) and (2.6) into the BRS identity,     
we obtain the following six sum rules:
\begin{subequations}
\begin{eqnarray}
  -2 \gamma^2 \,
  \Big\{ \widetilde{F}_{3,\tau}(s,t) - i\, \widetilde{F}_{4,\tau}(s,t) \Big\}
  + 4 \delta^2\, \widetilde{F}_{8,\tau}(s,t) +  \widetilde{F}_{13,\tau}(s,t) 
   &=& C_{\rm mod}^{BRS}  {H}_{1,\tau}(s,t) \;, \;\;
\label{eq-expbrs1a}
\\  
  - \widetilde{F}_{1,\tau}(s,t) 
  + 2 \gamma^2\, \widetilde{F}_{2,\tau}(s,t) 
  + \frac{1}{2} \widetilde{F}_{3,\tau}(s,t) 
  + \frac{i}{2} \widetilde{F}_{4,\tau}(s,t) 
  + \widetilde{F}_{14,\tau}(s,t) 
   &=& C_{\rm mod}^{BRS} {H}_{2,\tau}(s,t)\;, \;\;
\label{eq-expbrs1b}
\\ 
  - \frac{1}{2} \widetilde{F}_{5,\tau}(s,t) 
  - \frac{i}{2} \widetilde{F}_{6,\tau}(s,t)
  - \frac{\tau}{2} \widetilde{F}_{8,\tau}(s,t)
  + 2 \delta^2\, \widetilde{F}_{9,\tau}(s,t)
  + \widetilde{F}_{15,\tau}(s,t) 
   &=& C_{\rm mod}^{BRS} {H}_{3,\tau}(s,t)\;, \;\;
\label{eq-expbrs1c}
\end{eqnarray}
\end{subequations}
where
\begin{eqnarray}
  \gamma^2 = \frac{s}{4 m_W^2} ,\;\;\; 
  \delta^2 = \frac{s + 2 t - 2 m_W^2}{4 m_W^2} . 
\end{eqnarray}
Among the 18 physical form factors ($\widetilde{F}_{1,\tau}$ 
through $\widetilde{F}_{9,\tau}$ for $\tau=\pm 1$),  
all but the two $CP$-violating form factors $\widetilde{F}_{7,\tau}$ 
($\tau=\pm$) appear in the sum rules.  
The form factors $\widetilde{F}_{7,\tau}$ should be tested 
by other means. 
We find that the chargino and neutralino contributions 
to $\widetilde{F}_{7,\tau}$ are zero at the one-loop order.   
The remaining 16 physical form factors are tested by the sum rules 
(\ref{eq-expbrs1a})-(\ref{eq-expbrs1c}), where 
$\widetilde{F}_{13,\tau}$ through $\widetilde{F}_{15,\tau}$ 
are obtained from the $\eeww$ amplitude, and 
${H}_{1,\tau}$ through ${H}_{3,\tau}$ 
from the $e^+e^- \rightarrow \omega^+ W^-$ amplitude.
This extra effort is worthwhile because the test is very powerful;
each form factor has its own complicated dependence on $s$ and $t$.

We apply the BRS sum rules also for testing the numerical program. 
For this purpose, we have formulated the BRS sum rules to hold exactly 
for the one-loop form factors.  
Both sides of the six BRS sum rules should then agree within
the expected accuracy of the numerical computation.
We have confirmed that all six sum 
rules~(\ref{eq-expbrs1a})-(\ref{eq-expbrs1c}) hold to 
better than 11-digit accuracy at $e^+e^-$ collision energies 
$\sqrt{s}$ at 200, 500, and 1000 GeV.  
In the evaluation of the scalar one-loop integral functions, 
we have partly used the Fortran FF package~\cite{ff}.

\subsection{Decoupling limit}

\begin{figure}[t]
\begin{center}
\includegraphics*[width=8.5cm]{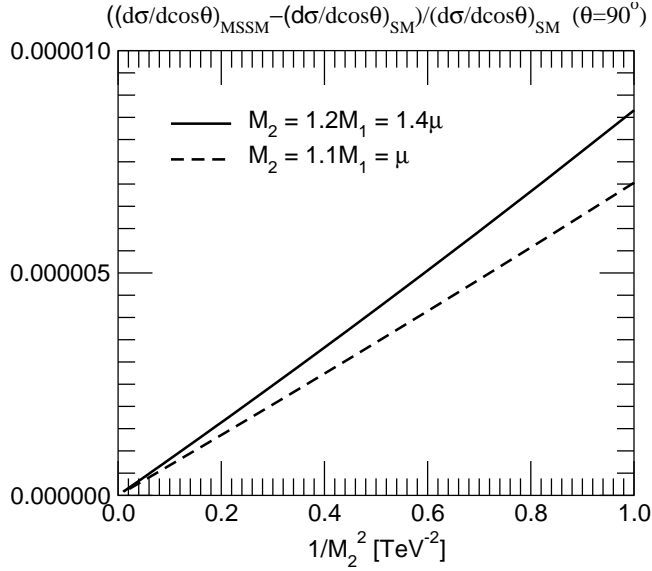} 
\caption{The test of decoupling of the chargino and neutralino 
contributions. The deviation of the helicity summed cross section from
the SM value versus 1/$M_2^2$ is shown at $\sqrt{s}=200$ GeV and  
the scattering angle $\theta = 90^{\circ}$, where $M_2$ is the
gaugino mass. The solid line is for 
$M_2$ = $1.2M_1$ = 1.4$\mu$, and the dashed line is for
$M_2$ = $1.1M_1$ = $\mu$. }
\label{fig:decoup}
\end{center}
\vspace*{0.5cm} 
\end{figure}

\hspace*{12pt}
The one-loop effects of the SUSY particles should decouple  
from the low energy observable in the large mass limit. 
The theory should then become effectively the SM. 
In the $\overline{\rm MS}$ scheme, perturbation expansion is 
performed by the $\msbar$ couplings of the MSSM, so that 
it is nontrivial to see the above statement of the decoupling clearly.
In order to show the decoupling openly, we use the 
$\overline{\rm MS}$ couplings of the SM as the expansion parameter 
of the perturbation theory.  
This is clearly the most convenient scheme below the 
SUSY particle threshold.
We adopt this scheme even above
the threshold,  because the difference
from the results in the $\overline{\rm MS}$
is found to be numerically very
small~\cite{eeww_sf}
as long as the logarithms of the ratios
$s/m_{\rm SUSY}^2$ are not too large.

In order to obtain a perturbative expression in terms of the 
${\overline{\rm MS}}$ couplings of the SM, we insert 
the expansion (3.1): 
\begin{subequations}
\begin{eqnarray}
\hat{e}^2 (\mu_R^{}) &\!=\!& 
\hat{e}^2_{\rm SM} (\mu_R^{})\left\{1 \!+\! 
 \frac{4}{3} \left[ \log{\frac{\mu^2_R}{m^2_{\wt{\chi}^{-}_1}}} 
+\log{\frac{\mu^2_R}{m^2_{\wt{\chi}^{-}_2}}}
\right] \frac{\hat{e}^2_{\rm SM} (\mu_R^{})}{16\pi^2} \right\}
,\\
\hat{g}^2 (\mu_R^{}) &\!=\!& 
\hat{g}^2_{\rm SM} (\mu_R^{})\left\{1 \!+\!
 \frac{2}{3} \left[\left\{ (D_L)_{11}\!+\!(D_R)_{11} \right\}
           \log{\frac{\mu^2_R}{m^2_{\wt{\chi}^{-}_1}}} 
+\left\{ (D_L)_{22}\!+\!(D_R)_{22} \right\}
           \log{\frac{\mu^2_R}{m^2_{\wt{\chi}^{-}_2}}}
\right]  \frac{\hat{g}^2_{\rm SM}(\mu_R^{})}{16\pi^2} 
\right\}\!\!, \makebox[10mm]{}
\end{eqnarray}
\label{eq:expand}
\end{subequations}
in all the form factors, 
and we retain only terms up to 
${\cal O}(\hat{g}^4_{\rm SM})$.
Hereafter, we perform this procedure in all our calculations. 
As a result of the expansion by SM coupling,
there is exactly no renormalization point dependence 
in our calculation.

In the large mass limit for charginos and neutralinos, 
the one-loop amplitudes behave as
\begin{eqnarray}
   \delta {\cal M}^{\rm Ino-loop} 
  \sim  A + B \frac{s}{m_{\widetilde{\chi}}^2} 
+ {\cal O} \left( \frac{s^2}{m_{\widetilde{\chi}}^4} \right) . 
\label{dec}
\end{eqnarray}
In the original expression for the amplitudes in terms of the MSSM $\msbar$ 
couplings, the constant term $A$ remains nonzero because higher order 
terms of ${\cal O}(\hatg^6)$ do not cancel exactly.  
On the other hand, in our scheme in which such higher 
order terms are systematically eliminated in the analytic expressions, 
the term $A$ in Eq.~\eq{dec} is exactly zero, and the decoupling 
of the chargino and neutralino effects is made manifest.
This property of the {\it exact decoupling} in our scheme can be used 
for an excellent test of the one-loop calculation 
including the overall normalization factors 
such as the $W$-boson wave-function renormalization constants that 
are not tested by the BRS sum rules. 
Figure~\ref{fig:decoup} shows the chargino and neutralino contributions
in the helicity summed differential cross section as a 
function of 1/$M^2_2$ at $\sqrt{s}$ = 200GeV and the
scattering angle $\theta$ = 90$^{\circ}$, where $M_2$ is the 
gaugino mass. The solid line is for 
$M_2$ = $1.2M_1$ = 1.4$\mu$, and the dashed line is for
$M_2$ = $1.1M_1$ = $\mu$.
We can see that the helicity summed differential 
cross section in the MSSM becomes that of the SM in both cases at
large mass of the gaugino.

\section{Numerical evaluation of the chargino and neutralino  
effects on $\eeww$}
\cleqn

\hspace*{12pt}
Having tested the numerical program in the last section, 
we are ready to study the one-loop chargino and neutralino 
contribution to the $\eeww$ helicity amplitudes.  
We present here the results of the one-loop contributions 
to the helicity amplitudes
as a function of the Higgs mixing parameter $\mu$ as well as 
of the $e^+e^-$ collider energy $\sqrt{s}$.

In Secs. 5.1 to 5.3, we show the results for $CP$ conserving cases. 
The free parameters in the chargino and neutralino sectors are  
then the $\mu$ parameter (and its sign), 
the ratio of the vacuum expectation value $\tan\beta$, 
and the soft SUSY breaking gaugino masses $M_1$ and $M_2$ 
for $U(1)$ and $SU(2)$, respectively.    
For simplicity, we assume the relation 
$M_1 = 5 M_2 \hat{s}^2/3\hat{c}^2$ throughout this paper.  
The MSSM parameter sets (set 1 to set 7) 
that we adopt for the figures showing 
the $\mu$ dependences are summarized in Table~\ref{tab:set2}. 
The two signs of the $\mu$ parameter, the two extreme values 
of $\tan\beta$ (3 and 50), and four values of the lightest 
chargino mass ($m_{\widetilde{\chi}_1^-}=110, 130,150$, and $170$ GeV) 
are examined. 
The $\sqrt{s}$ dependences of the helicity amplitudes 
are studied in the MSSM parameter sets (set A to set E) 
given in Table~\ref{tab:set1}.     
All five cases are for 
$m_{\widetilde{\chi}_1^-}=110$ GeV, $\tan\beta=3$, and ${\rm sgn}(\mu)=+$. 
They have different values of the ratio $\mu/M_2$. 
The last case (set E) has $CP$-violating phases 
$\varphi_1$ and $\varphi_\mu$ of $M_1$ and $M_\mu$, respectively. 
In Sec. 5.4, we discuss the case of 
nonzero $\varphi_1$ and $\varphi_\mu$ in set E of Table~\ref{tab:set1}. 

\begin{table}[t]
\begin{center}
\vspace{1cm}
\begin{tabular}{l|rrrrrrrr}
   & 1 & 2 & 3 & 4 & 5 & 6 & 7 \\ \hline
\multicolumn{7}{l}{Parameter}          \\ \hline
sgn($\mu$) &   + & $-$ &  + & $-$ &+&+&+ \\
$\tan\beta$ &   3 &   3 &  50 &  50 &3&3&3\\ 
$m_{\wt{\chi}^-_1}$ (GeV) & 110 & 110 & 110 & 110&130&150&170 \\
\hline\hline
\end{tabular}
\end{center}
\caption{The parameter sets for figures of showing $|\mu |$ 
         dependence. Sets 1$-$4 are for Fig.~3(a), Fig.~4(a), and 
Fig.~6(a).  Sets 5$-$7 are used in Fig.~5. }
\label{tab:set2}
\end{table}

\begin{table}[t]
\begin{center}
\begin{tabular}{l|rrrrr}
   & A & B & C & D & E\\ \hline
\multicolumn{5}{l}{Parameter}           \\ \hline
$\mu$ (GeV)     & +120 & +145 & +400 & +1000 & +130 \\
$M_2$ (GeV)     & 541 & 242 & 125 & 115  & 158 \\
$\tan\beta$&   3 &   3 &   3 &    3 &   3 \\
$\varphi_{1}$  &0&   0 &   0 &    0 & $\frac{2}{3}\pi$ \\
$\varphi_{\mu}$&0&   0 &   0 &    0 & $\frac{2}{3}\pi$ \\
\hline  
\multicolumn{5}{l}{Mass spectra (GeV)} \\ \hline
$m_{\widetilde{\chi}^-_1}^{}$  & 110 & 110 & 110 & 110 & 110\\
$m_{\widetilde{\chi}^-_2}^{}$  & 555 & 283 & 420 &1007 & 207\\
$m_{\widetilde{\chi}^0_1}^{}$  &  99 &  81 &  60 &  57 &  75\\
$m_{\widetilde{\chi}^0_2}^{}$  & 123 & 150 & 111 & 110 & 105\\
$m_{\widetilde{\chi}^0_3}^{}$  & 285 & 150 & 403 &1002 & 154\\
$m_{\widetilde{\chi}^0_4}^{}$  & 555 & 285 & 422 &1007 & 205\\
\hline\hline
\end{tabular}
\end{center}
\caption{The parameters and the mass spectrum for the figures of 
$\sqrt{s}$ dependence. Sets A$-$D  are used 
in Figs.~3(b), 4(b), and 6(b).
Set E is the $CP$-violating case and used in Figs.~7(a) and 7(b). 
The lightest chargino mass is fixed to be 110 GeV for all sets. 
}
\label{tab:set1}
\end{table}

We show the one-loop contributions of charginos and neutralinos 
to each helicity amplitude in the form 
\begin{eqnarray}
 \frac{{M_{\tau}^{\lambda \overline{\lambda}}}_{\rm MSSM}
       - {M_{\tau}^{\lambda \overline{\lambda}}}_{\rm SM}}
  {|{M_{\tau}^{\lambda \overline{\lambda}}}_{\rm SM}|}, 
\end{eqnarray}
where ${M_{\tau}^{\lambda \overline{\lambda}}}_{\rm MSSM}$ are the helicity 
amplitudes of the MSSM in which only one-loop chargino and neutralino 
contributions are considered, and  
${M_{\tau}^{\lambda \overline{\lambda}}}_{\rm SM}$ are those of the SM. 
From this expression, not only the ratio of the SUSY 
contributions to the SM amplitude
but also its sign (for the real and imaginary parts)
can be inferred. 

The magnitude and the sign of all the SM amplitudes 
at the scattering angle 
$\theta=90^{\circ}$ 
are shown in Fig.~2(a) versus the $e^+e^-$ collision energy $\sqrt{s}$.
Among the tree-level helicity amplitudes, 
$M^{+-}_{\tau=-1}$,  $M^{-+}_{\tau=-1}$,  
and $M^{00}_{\tau=\mp 1}$ are significant for all energies.
The other helicity amplitudes are reduced as $\sqrt{s}$ grows; i.e., 
$M^{0\pm}_{\tau=\pm}$ and $M^{\pm 0}_{\tau=\pm}$ 
($M^{++}_{\tau=\pm}$ and $M^{--}_{\tau=\pm}$) behave as 
$1/\sqrt{s}$ ($1/s$)~\cite{eeww_sf}. 
For $\sqrt{s} < 274$ GeV, $M^{0+}_{\tau=-1}$ and $M^{-0}_{\tau=-1}$ 
are larger than $M^{00}_{\tau=-1}$ at $\cos\theta=0$.
In the following (Secs.~5.1, 5.2, and 5.3), 
we show the one-loop effects on the helicity amplitudes of 
$M^{\pm\mp}_{\tau=-1}$, $M^{00}_{\tau=\mp 1}$, and 
$M^{0+(-0)}_{\tau=-1}$, respectively, in the $CP$ conserving cases. 
In Sec.~5.4, we examine the loop-induced $CP$-violating effects on 
the vertex form factors $f_4^Z$ and $f_6^V$ ($V=\gamma$ and $Z$) 
and show their contributions to the helicity amplitudes 
$M^{0\pm (\mp 0)}_{\tau=-1}$.

In Fig.~2(b), for completeness, the corresponding cross sections 
integrated for $|\cos\theta| < 0.8$ are shown for each helicity set.   
The results for the helicity summed total cross section are also shown. 

\begin{figure}[t]
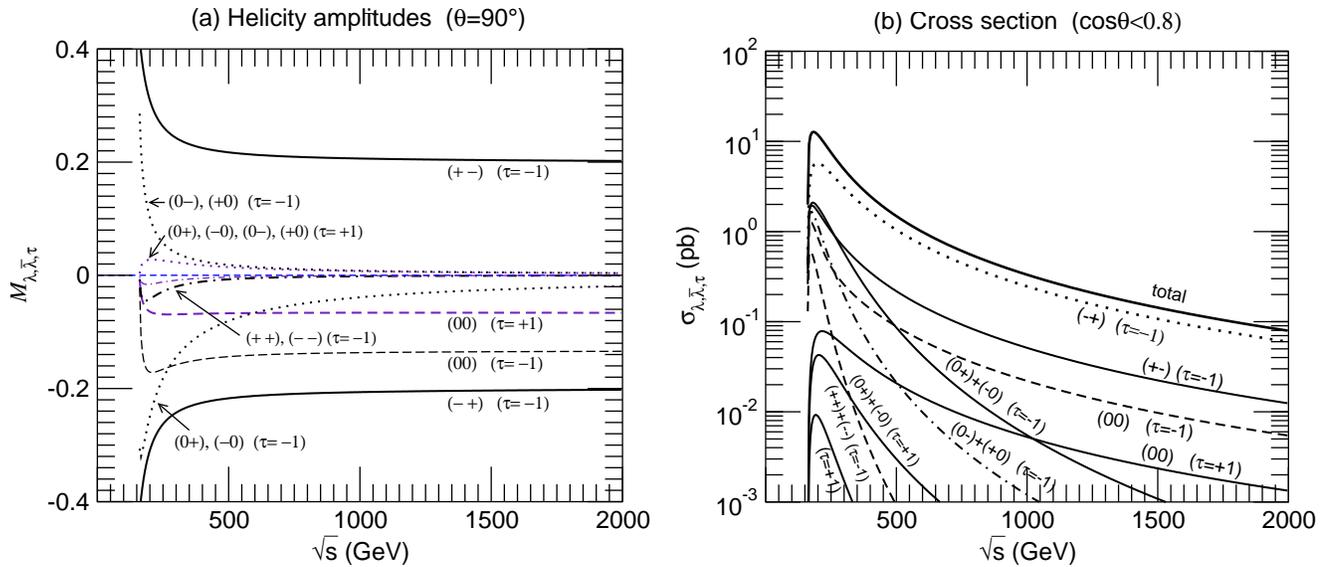

\begin{center}
$\begin{array}{cc}
\label{fig:amp_tree}
\includegraphics*[width=8.5cm]{2a.epsi}  &
\includegraphics*[width=8.5cm]{2b.epsi} 
\label{fig:cross_total}
\end{array}$
\end{center}
\caption{{\bf (a)} 
The tree-level helicity amplitudes of $\eeww$ for each set of  
$\lambda,\overline{\lambda}$, and $\tau$ at the scattering angle $90^\circ$. 
{\bf (b)} The total cross section with $|\cos\theta| <0.8$ for 
each helicity set of $\lambda,\overline{\lambda}$, and $\tau$.}
\end{figure}


\subsection{The chargino and neutralino contributions 
            to $M_{\tau}^{\pm\mp}$}

\hspace*{12pt}
The helicity amplitudes $M^{+-}_{\tau=-1}$ and $M^{-+}_{\tau=-1}$  
are the largest of all the helicity amplitudes 
at large scattering angles. 
At the tree level, only the $t$-channel neutrino-exchange diagram 
contributes to the $(+-)$ and $(-+)$ amplitudes.
The one-loop contribution of charginos and neutralinos 
to these helicity amplitudes comes only from the $W$-boson 
wave-function renormalization factor. 
Therefore, the one-loop effects are essentially independent 
of the $e^+e^-$ collision energy $\sqrt{s}$ and 
the scattering angle $\theta$,   
and they are determined by a logarithmic function of the masses of 
charginos, neutralinos, and the $W$ boson. 

\begin{figure}[p]
\begin{center}
$\begin{array}{cc}
\includegraphics*[width=7.5cm]{3a.epsi} & 
\includegraphics*[width=7.5cm]{3b.epsi}
\end{array}$
\end{center}
\label{fig:amp+-}
\caption{The dependence on 
(a) $|\mu |$ and (b) $\sqrt{s}$ of the 
chargino and neutralino one-loop contributions to $M_{\tau=-1}^{+-}$ 
are shown at $\theta = 90^{\circ}$ 
for $m_{\widetilde{\chi}_1^-}=110$ GeV. 
In Fig. (a), parameters of set 1 to set 4
in Table~\ref{tab:set2} are used. 
The $e^+e^-$ collision energy $\sqrt{s}$ is 220 GeV.
In Fig. (b), parameters of set A to set D of
Table~\ref{tab:set1} are used.
}
\begin{center}
$\begin{array}{cc}
\includegraphics*[width=7.5cm]{4a.epsi}& 
\includegraphics*[width=7.5cm]{4b.epsi}\\
\end{array}$
\label{fig:amp00}
\end{center}
\caption{
The dependence on 
(a) $|\mu |$ and (b) $\sqrt{s}$ of the 
chargino and neutralino one-loop contributions to $M_{\tau=\pm}^{00}$ 
are shown at $\theta = 90^{\circ}$ 
for $m_{\widetilde{\chi}_1^-}=110$ GeV. 
In Fig. (a), parameters of set 1 to set 4
in Table~\ref{tab:set2} are used. 
The $e^+e^-$ collision energy $\sqrt{s}$ is 220 GeV.
In Fig. (b), parameters of set A to set D of
Table~\ref{tab:set1} are used.
}
\end{figure}

In Fig.~3(a), we show the $|\mu |$ dependence in 
$M^{+-}_{\tau=-1}$ at the scattering angle $\theta = 90^\circ$.
The input parameters are summarized in Table~\ref{tab:set2}.
The mass of the lightest chargino is fixed to be 110 GeV 
for all cases, so that the ratio $M_2/\mu$ is a constant 
for each fixed value of $\tan\beta$ and $M_1$. 
The $e^+e^-$ collision energy $\sqrt{s}$ is set to 
be at the threshold of the lightest chargino pair 
production; i.e., $\sqrt{s}=220$ GeV.
In the large $|\mu|$ region, the lightest chargino 
is Wino-like, i.e., the mass comes from $M_2^{}$.   
We confirmed numerically that in the limit of $\mu \to \infty$, 
the deviation becomes constant for $\mu$. 
This reflects the fact that the lightest chargino is 
purely Wino-like, and the effect of the Higgsino 
decouples from the one-loop helicity amplitudes 
$M^{+-}_{\tau=-1}$ and $M^{-+}_{\tau=-1}$. 
The deviation at $|\mu | =1000$ GeV is about 0.08$\%$ for set 1$-$4.
For smaller $|\mu|$ values, $M_2$ becomes larger so that 
the lightest chargino contribution becomes smaller because of 
decoupling.   
On the contrary, for $|\mu|$ around 110 GeV, 
the lightest chargino is Higgsino-like, i.e., 
$m_{\widetilde{\chi}_1^-} \sim |\mu|$.  For smaller 
values of $|\mu|$, a larger Higgsino-like contribution appears. 
The Wino-like contribution, which is enhanced for the large $|\mu|$ 
region, and the Higgsino-like contribution, which is substantial 
for small $\mu$ values have the same sign.   
Therefore, the deviation reaches its minimum 
at $|\mu | = 140$ GeV for set 1 and $|\mu | = 124$ GeV 
for set 3 and set 4.
For set 2, the deviation monotonically increases, 
because in this case $M_2$ is similar to or less than $\mu$ 
even around $\mu=110$ GeV, so that the Higgsino contribution 
is smaller than the Wino contribution. 
The results for set 3 and set 4 are similar, because the mass 
eigenstates of the chargino and neutralino fields are common
between set 3 and set 4 in the limit of large $\tan\beta$. 

In Fig.~3(b), $M^{+-}_{\tau=-1}$ is shown as a 
function of the $e^+e^-$ collider energy 
$\sqrt{s}$ at $\theta = 90^\circ$ for the  parameters of 
set A to set D in Table~\ref{tab:set1}.  The lightest chargino mass
is again fixed to be 110 GeV, and $\mu$ is assumed to be positive 
and $120$, $145$, $400$, and $1000$ GeV for set A, set B, set C and set D, 
respectively. 
The corrections are insensitive to $\sqrt{s}$, because 
there is no Feynman diagram of one-loop charginos and neutralinos 
which contribute to $M^{+-}_{\tau=-1}$.
As we do not include the SM one-loop diagrams, 
the renormalization scale $\mu_R^{}$ dependence 
which comes from the SM running effect in the $\overline{\rm MS}$ 
couplings remains in our calculation. 
By setting $\mu_R^{}$ to be $\sqrt{s}$,  
an artificial tiny $\ln s$ dependence appears 
in $M^{+-}_{\tau=-1}$. 

The magnitude of the chargino and neutralino contributions to 
$(+-)$ is small.

\subsection{The chargino and neutralino contributions to 
$M_\tau^{00}$}

\hspace*{12pt}
The one-loop corrections to the trilinear gauge couplings 
are expected to affect the helicity amplitude $M^{00}_{\tau}$ 
significantly, because $M^{00}_{\tau}$ includes 
contributions from 
s-channel $Z$ boson and photon exchange diagrams. 

\begin{figure}[p]
\begin{center}
\includegraphics*[width=8.5cm]{5.epsi} 
\label{fig:amp00cino}
\end{center}
\caption{The $|\mu |$ dependences of the chargino and neutralino
one-loop contributions to $M_\tau^{00}$ 
are shown for $m_{\widetilde{\chi}_1^-}=110$, 
130, 150, and 170 GeV .
The $e^+e^-$ collider energy $\sqrt{s}$ is taken to be 
$2 m_{\widetilde{\chi}_1^-}$. 
}
\begin{center}
$\begin{array}{cc}
\includegraphics*[width=8.5cm]{6a.epsi} &
\includegraphics*[width=8.5cm]{6b.epsi}
\end{array}$
\end{center}
\caption{
The dependence on 
(a) $|\mu |$ and (b) $\sqrt{s}$ of the 
chargino and neutralino one-loop contributions to $M_{\tau=-1}^{0+}$ 
are shown at $\theta = 90^{\circ}$ 
for $m_{\widetilde{\chi}_1^-}=110$ GeV. 
In Fig. (a), parameters of set 1 to set 4
in Table~\ref{tab:set2} are used. 
The $e^+e^-$ collision energy $\sqrt{s}$ is 220 GeV.
In Fig. (b), parameters of set A to set D of
Table~\ref{tab:set1} are used.
}
\label{fig:amp0+}
\end{figure}

In Fig.~4(a), we show the effects of charginos and neutralinos 
on $M^{00}_{\tau}$  ($\tau=\pm$) at $\theta=90^\circ$ 
and at the threshold of the lightest chargino-pair production 
($\sqrt{s}=220$ GeV)
when $m_{\widetilde{\chi}_1^-}=110$ GeV.
The four curves each for $\tau=-1$ and $+1$ correspond to 
the parameter sets (set 1 to set 4) in Table~2.
Like $M^{+-}_{\tau}$, the Wino effects dominate 
in the large $|\mu|$ region, while the Higgsino 
contributes for the small $|\mu|$ region. 
The effects grow at large values of $|\mu|$ for $\tau=-1$ for all 
cases, up to about 0.7\% at $|\mu|=1000$ GeV, whereas 
they remain small for $\tau=+1$, at around the $-0.1$\% level.

In Fig.~4(b), the one-loop contributions of 
charginos and neutralinos to $M^{00}_{\tau=\pm 1}$ are shown 
as a function of $\sqrt{s}$ at $\theta = 90^\circ$ 
for $\tan\beta$=3 and $\mu>0$. 
The four sets of parameters (sets A to D) correspond to 
the different values of $|\mu|$ as listed in Table~\ref{tab:set1}.
Let us see the $\tau=-1$ amplitudes first. 
Sharp peaks can be seen for each curve, which 
correspond to the thresholds of pair production of 
the lightest charginos and the two lightest neutralinos.
The deviation at the threshold ($\sqrt{s}$=220 GeV) 
can reach $0.36$\% for set A, $0.19$\% for set B, $0.55$\% 
for set C, and $0.72$\% for set D.
For $\tau=+1$, the chargino and neutralino corrections from the SM
are negative, and hence they interfere constructively with 
the negative SM amplitude (see Fig.~2(a)). 
The deviations from the SM prediction at $\sqrt{s}$=220 GeV 
are $0.0\%$ for set A, $-0.15\%$ for set B, and $-0.08\%$ for 
set C and set D. 
The deviations from the SM are $-0.31\%$ at the first threshold of 
neutralino production and $-0.33\%$ at the second threshold of
neutralino production for set B.
Notice that the tree level amplitude of $M^{00}_{\tau = -1}$
is already twice that of $M^{00}_{\tau = +1}$, so that 
the one-loop chargino and neutralino contributions to 
$M^{00}_{\tau = -1}$ at the threshold of light chargino pair production 
are much larger than those to the $M^{00}_{\tau = +1}$ amplitude.

Finally, in Fig.~5, we show the corrections for different values of 
the lightest chargino mass as a function of $|\mu|$. 
The four curves in the figure correspond to set 1, 5, 6, and 7 
of Table 2, in which the mass of the lightest charginos are 
set to be 110, 130, 150, and 170 GeV, respectively.
In the large $|\mu|$ region where the lightest chargino is 
Wino-like, the deviation from the SM value is reduced
as $m_{\widetilde{\chi}_1^-}$ grows. 
The deviation at $|\mu|=1000$ GeV changes from 0.72\% to 0.64 \% 
when $m_{\widetilde{\chi}_1^-}$ is taken to be
110 GeV (set~1) and 170 GeV (set~7).
For smaller values of $|\mu|$ where the lightest chargino 
is Higgsino-like, the value of the biggest Higgsino contribution  
at the threshold of the lightest chargino pair production 
is almost the same for all cases and is about 0.4\%.

\subsection{The chargino and neutralino contributions to 
$M_{\tau=-1}^{0+}$ and $M_{\tau=-1}^{-0}$}

\hspace*{12pt}
As already mentioned, the tree-level helicity amplitudes of 
$M_{\tau}^{0+}$ and $M_{\tau}^{-0}$ behave as ${\cal O}(1/\sqrt{s})$, 
so that they are substantial only at relatively low energies. 

We here present results for the chargino and neutralino one-loop 
contributions to $M_{\tau=-1}^{0+}$ in Figs.~6(a) and 6(b).
We find similar characteristics to the 
corrections to $M_{\tau=-1}^{00}$, whose details have already been 
discussed.
The magnitude of the deviation from the SM amplitudes 
is no larger than that of $M_{\tau=-1}^{00}$ 
for each parameter set at the threshold of light chargino pair production.

\subsection{The $CP$-violating effects}

\hspace*{12pt}
In the general MSSM, there are new $CP$-violating phases.  
$CP$-violating form factors for the $WW\gamma$ and $WWZ$ vertices 
($f_4^V$, $f_6^V$, and $f_7^V$ with $V=\gamma$ and $Z$) 
can be induced beyond the tree level due to the SUSY particle loops. 

The $CP$-violating phases in the chargino and neutralino sectors 
arise from the $\mu$ parameter and the gaugino mass parameters  
$M_1$ and $M_2$. The other sector of the MSSM Lagrangian 
also includes $CP$-violating phases, such as in the gluino 
mass parameter $M_3$ and  the trilinear $A$ terms of sfermions. 
The experimental upper bounds on the electric 
dipole moments (EDM's) of electrons and neutrons 
provide very severe constraints on those  
$CP$-violating phases~\cite{cp1}.
It has been found that 
internal cancellation of the phases in the EDM's may still allow
for relatively large $CP$-violating phases~\cite{cp2}. 
Large $CP$-violating phases in the chargino and neutralino sectors 
are possible without contradicting the EDM constraint, 
if the parameters for sleptons and squarks of the first generation 
are adjusted.  
As we can take the phase of $M_2$ to be 0 by rephasing, 
the dependence on the phase of $\mu$ ($\varphi_\mu$) and that of 
$M_1$ ($\varphi_1$) is examined in this paper.  
Here, we study the case in which the large 
$CP$-violating effects on the $WW\gamma$ and $WWZ$ coupling appear, 
and examine the deviation in the helicity amplitudes 
from the $CP$ conserving case. 
We note that our numerical results in this section are consistent with 
the result previously obtained by Kitahara et al.~\cite{oshimo}.

Among the 18 physical form factors of $\eeww$ (see Eq.~(2,2)), 
$F_4$, $F_6$, and $F_7$ have the $CP$-odd property.  
Chargino and neutralino triangle type diagrams for  
the triple gauge vertices $VWW$ ($V=\gamma$ or $Z$) 
contribute to the $CP$-violating form factors $f^{Z (1)}_4$, 
$f^{\gamma (1)}_6$, and $f^{Z (1)}_6$ at one loop. 
We note that the chargino and neutralino loop diagrams do not 
contribute to $f^{V (1)}_7$, so that $F_7$ is zero   
(The relation between the form factors $F_{i,\tau}^{}$ 
of the $\eeww$ amplitude and the form factors 
$f_{i}^{V}$ of the $WWV$ vertices).

\begin{figure}[t]
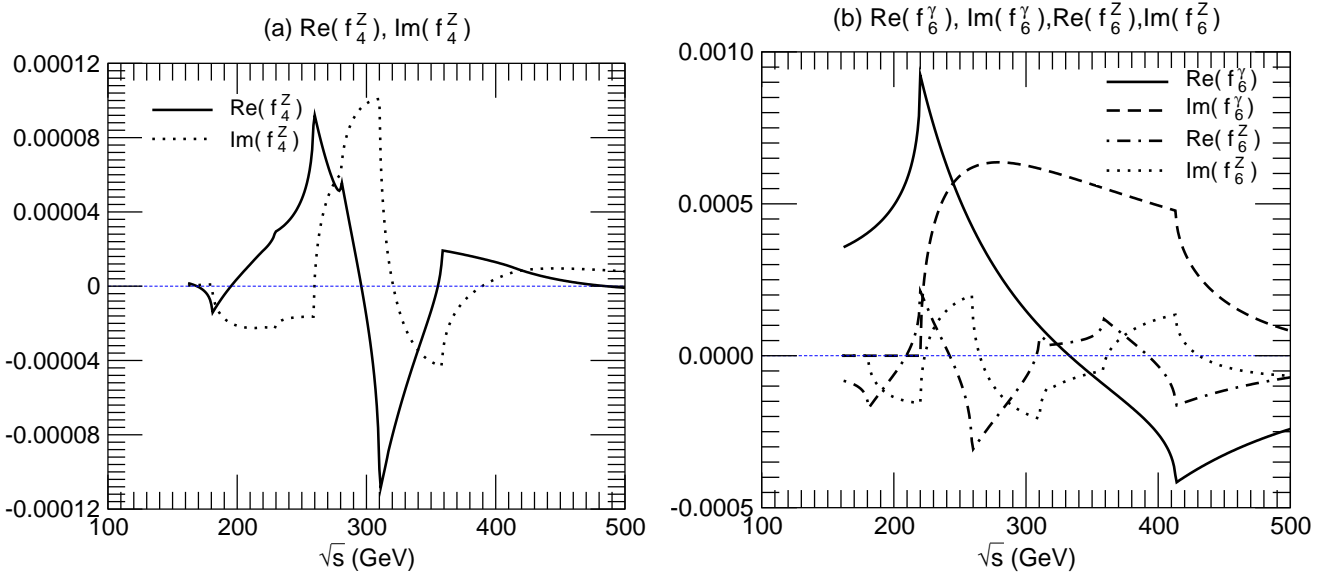

\begin{center}
$\begin{array}{cc}
\includegraphics*[width=8.5cm]{7a.epsi} &
\includegraphics*[width=8.5cm]{7b.epsi} 
\end{array}$
\end{center}
\caption{The $\sqrt{s}$ dependences of $f^\gamma_6$ and $f^Z_6$
are shown for $\varphi_1=\varphi_\mu$ = 2$\pi$/3. 
The lightest chargino mass is fixed as 110 GeV and 
$\mu$=130 GeV.
The parameters of set E in Table~3 are used.
}
\label{fig:phase_6g6z_rs}
\vspace*{1cm}
\end{figure}

\begin{figure}[p]
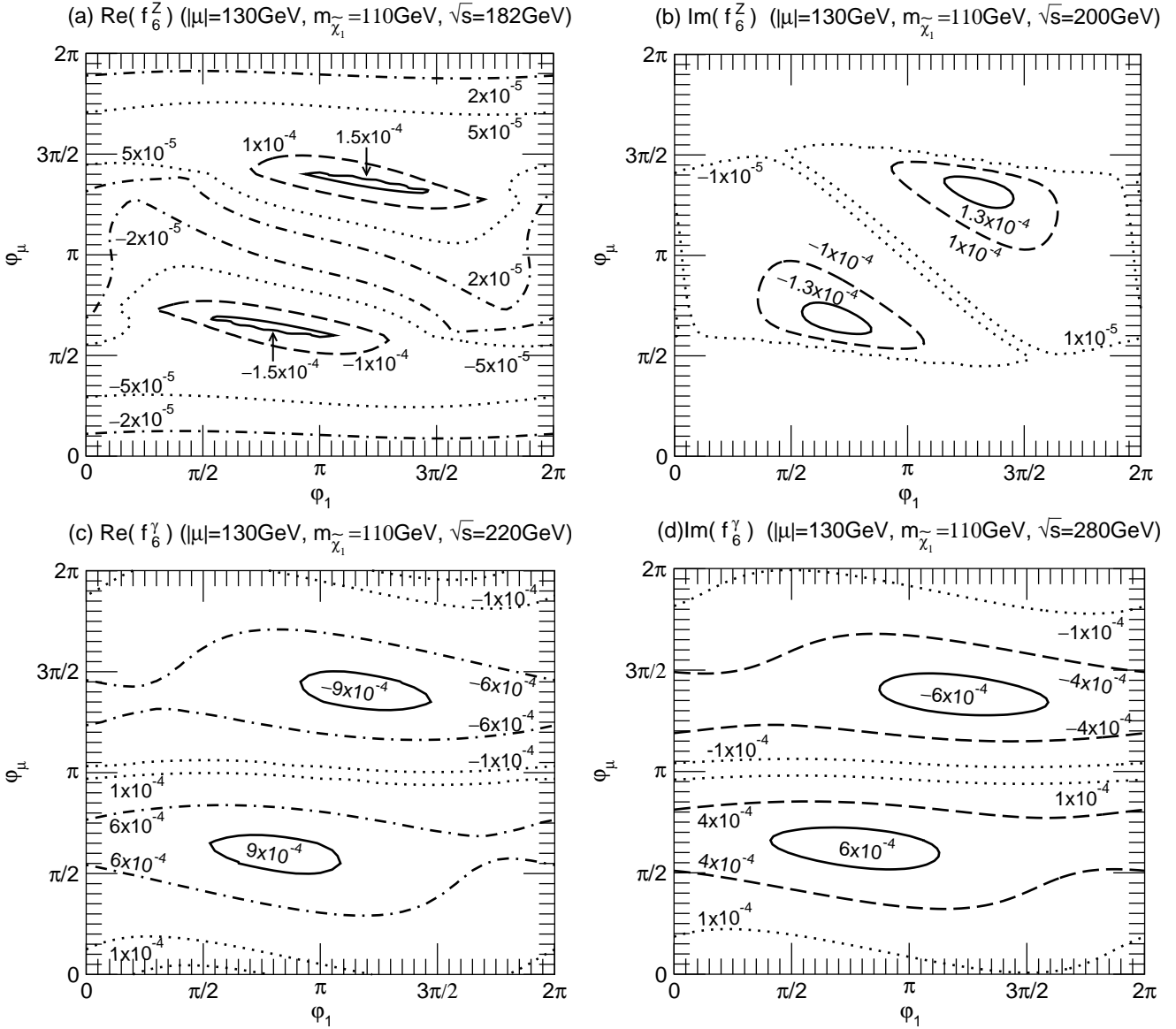

\begin{center}
$\begin{array}{cc}
\includegraphics*[width=8.5cm]{8a.epsi} &
\includegraphics*[width=8.5cm]{8b.epsi} \\
\includegraphics*[width=8.5cm]{8c.epsi} &
\includegraphics*[width=8.5cm]{8d.epsi} 
\end{array}$
\end{center}
\caption{The contour plots of $f^\gamma_6$ and $f^Z_6$ 
are shown in the $\varphi_1$-$\varphi_\mu$ plane.
The $e^+e^-$ collision energy $\sqrt{s}$ is taken to be 
182, 200, 220, and 280 GeV for Re($f^Z_6$),
Im($f^Z_6$), Re($f^\gamma_6$), and Im($f^\gamma_6$), respectively.
The lightest chargino mass $m_{\widetilde{\chi}_1^-}$ 
is fixed as 110 GeV, and $|\mu|$ is taken to be 130 GeV.
}
\label{fig:phase_6g6z}
\vspace*{1cm}
\end{figure}

\begin{figure}[p]
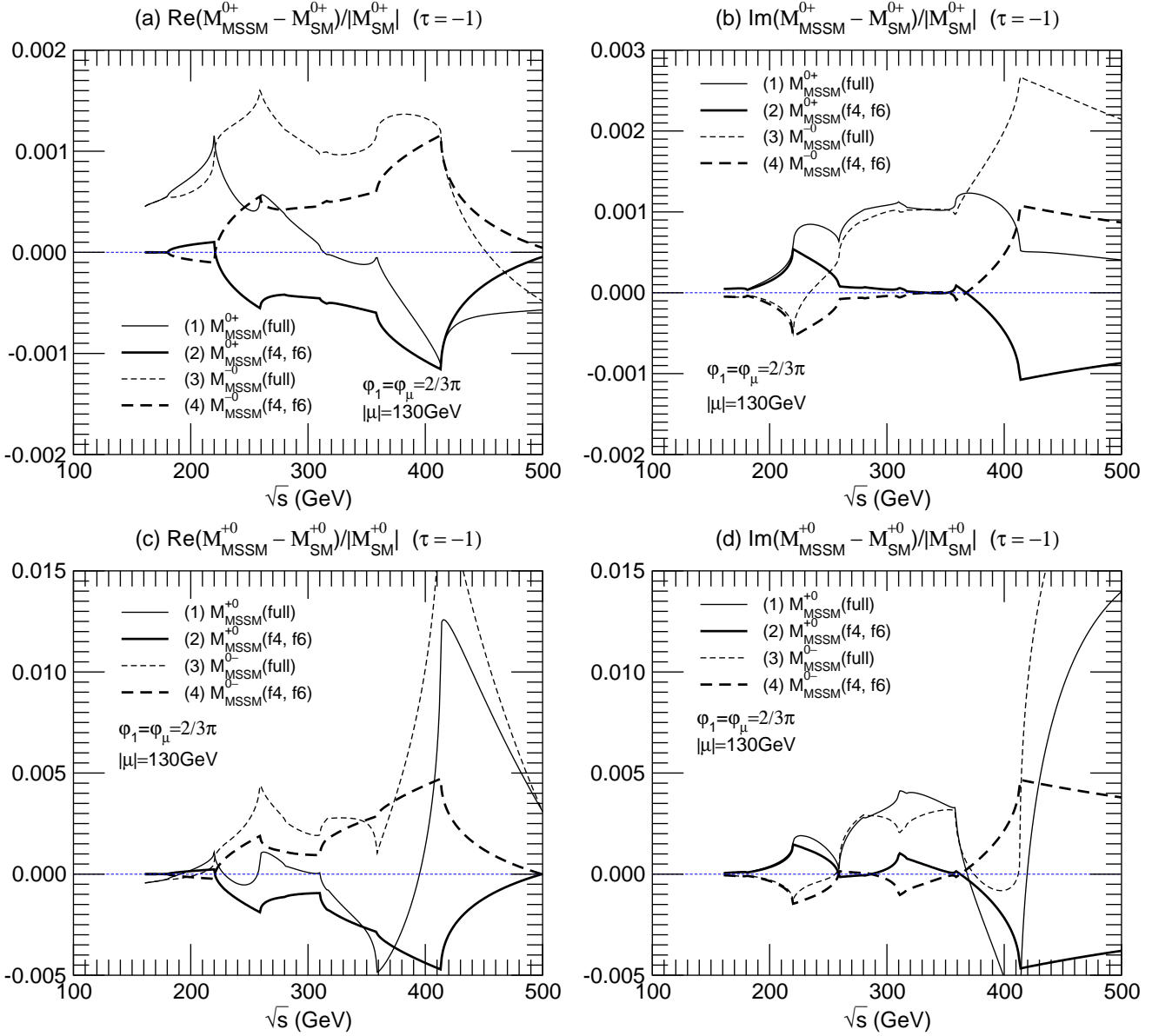

\begin{center}
$\begin{array}{cc}
\includegraphics*[width=8.5cm]{9a.epsi} &
\includegraphics*[width=8.5cm]{9b.epsi} \\
\includegraphics*[width=8.5cm]{9c.epsi} &
\includegraphics*[width=8.5cm]{9d.epsi} 
\end{array}$
\end{center}
\caption{The $\sqrt{s}$ dependences of the chargino and neutralino 
contributions to the real part and the imaginary part 
of $M_{\tau=-1}^{0+}$, 
$M_{\tau=-1}^{-0}$, 
$M_{\tau=-1}^{+0}$, and 
$M_{\tau=-1}^{0-}$ are shown
at $\theta=90^{\circ}$ for $m_{\widetilde{\chi}_1^-}=110$ GeV. 
The $CP$ phases $\varphi_1$ and $\varphi_{\mu}$ are
set to be 2/3$\pi$. $|\mu|$ is fixed to be 130 GeV. 
The parameters of set E in Table~3 is used.
The helicity amplitude $M_{{\rm MSSM}}$(full) contains 
contributions from all form factors, while 
$M_{{\rm MSSM}}$(f4, f6) includes only 
the contributions from the form factors $F_4$ and $F_6$. 
}
\label{fig:M0+_f4f6vsfull}
\vspace*{1cm}
\end{figure}

In Fig.~7(a), the real part (solid curve) and the 
imaginary part  (dotted curve) of $f^{Z}_4$ are shown 
as a function of $\sqrt{s}$ 
for the parameters of set E in Table~\ref{tab:set1}. 
The lightest chargino mass is fixed to be $110$ GeV,  
$|\mu|$ is 130 GeV, and $\tan\beta=3$.
The $CP$-violating phases $\varphi_1$ and 
$\varphi_\mu$ are taken to be $2\pi/3$.  
The threshold of the neutralino pair production 
for $\widetilde{\chi}^0_1\widetilde{\chi}^0_2$, 
$\widetilde{\chi}^0_1\widetilde{\chi}^0_3$, 
$\widetilde{\chi}^0_2\widetilde{\chi}^0_3$, 
$\widetilde{\chi}^0_1\widetilde{\chi}^0_4$, 
$\widetilde{\chi}^0_2\widetilde{\chi}^0_4$, 
or $\widetilde{\chi}^0_3\widetilde{\chi}^0_4$ 
is located at $\sqrt{s}=180$, $229$, $259$, $280$, or 
$310$ GeV, respectively. 
The real part has a peak at each threshold, while the 
imaginary part shows a rapid $s$-wave rise 
above the threshold. 
The magnitude of ${\rm Re}(f^Z_4)$ and ${\rm Im}(f^Z_4)$ 
is small and is at most $10^{-4}$. 

The $\sqrt{s}$ dependences of the 
real part and the imaginary part of $f^{\gamma}_6$ and $f^{Z}_6$ 
are shown in Fig.~7(b) for the same parameter choice 
as in Fig.~7(a); i.e., set E of Table~\ref{tab:set1}. 
The solid and dashed curves correspond to 
${\rm Re}(f^{\gamma}_6)$ and ${\rm Im}(f^{\gamma}_6)$, 
while the dot-dashed and dotted curves represent 
${\rm Re}(f^{Z}_6)$ and ${\rm Im}(f^{Z}_6)$, respectively. 
In the one-loop triangle diagrams for $\gamma W^+W^-$, 
only the chargino loops appear 
with the diagonal vertices 
$\gamma \widetilde{\chi}_i^-\widetilde{\chi}^+_i$ ($i=1,2$), 
so that the threshold effects for 
$\widetilde{\chi}_1^+\widetilde{\chi}_1^-$ 
and $\widetilde{\chi}_2^+\widetilde{\chi}_2^-$ 
pair production are seen at $220$ and $414$ GeV, respectively.
At the lightest chargino pair production threshold, 
the magnitude of ${\rm Re}(f_6^\gamma)$ can reach about $10^{-3}$. 
We note that at $\sqrt{s}=220$ GeV 
the maximal value of $f_6^{\gamma(1)}$ strongly depends 
on $\tan\beta$; it is 0.0017, 0.0013, or 0.00027 for 
$\tan\beta= 1, 2$, and $10$, respectively. 
For larger $\tan\beta$, smaller values of $f_6^{\gamma(1)}$ 
are obtained.
On the other hand, $f^{Z}_6$ shows more complicated behavior due 
to the threshold structure of neutralino pairs as well as chargino pairs. 
The magnitude of the real part is $-1.7~(+2.4, -3.2) \times 10^{-4}$ 
at the threshold of $\widetilde{\chi}^0_1\widetilde{\chi}^0_2$ 
($\widetilde{\chi}^+_1\widetilde{\chi}^-_1$, 
$\widetilde{\chi}^0_2\widetilde{\chi}^0_3$). 
The dotted line is the imaginary part of $f^{Z}_6$.
The magnitude of the imaginary part is as large as 
that of the real part.  

In Fig.~8, contour plots of 
(a) ${\rm Re}(f^{Z}_6)$, 
(b) ${\rm Im}(f^{Z}_6)$,
(c) ${\rm Re}(f^{\gamma}_6)$,
and 
(d) ${\rm Im}(f^{\gamma}_6)$ are shown 
in the $\varphi_1$-$\varphi_{\mu}$ plane. 
The lightest chargino mass is fixed as 110 GeV. 
We have chosen $|\mu|$ to be $130$ GeV, where relatively large values of 
$f^{\gamma}_6$ and $f^{Z}_6$ are observed. 
The relation $|M_1|$ = $|M_2| 5\hat{s}^2 / 3\hat{c}^2$ 
is used even though we allow $M_1/M_2$ to have an arbitrary phase 
$\varphi_1$.  
The value of $M_2$ is set so as to give 
the mass of the lightest chargino to be 110 GeV with $\tan\beta$=3. 
We also choose $\sqrt{s}$ to be 
182, 200, 220, and 280 GeV for the figures of 
Re$(f^Z_6)$, Im$(f^Z_6)$, Re$(f^\gamma_6)$, 
and Im$(f^\gamma_6)$, respectively,
where the form factors are relatively large (See Figs.~7(a) and 7(b)).
$\varphi_1$ and $\varphi_{\mu}$ are allowed to vary between 0 and $2\pi$.
As shown in the figures, 
both $f_6^Z$ and $f_6^\gamma$ take their maximum or minimum at around 
$\varphi_1$ =  $\varphi_{\mu}$ =$2\pi /3$ or $4\pi /3$. 
Re$(f_6^Z)$ (Im$(f_6^Z)$) can be 1.5~(1.3) $\times 10^{-4}$, while 
Re$(f_6^\gamma)$ (Im$(f_6^\gamma)$) can be larger than 
9.0~(6.0) $\times 10^{-4}$. 

In Figs.~9(a) to 9(d), we show the effects of nonzero 
$f_4^{Z}$ and $f_6^{Z,\gamma}$ on 
the helicity amplitude of $M^{0+}$, $M^{-0}$, $M^{+0}$ and $M^{0-}$
(see Eqs.~(\ref{m0+}) and (\ref{m+0})). 
In Fig.~9(a) (9(b)), the real (imaginary) part of the 
deviation in $M^{0+}$, and $M^{-0}$ with $\tau=-1$ 
from the SM prediction is shown 
for (1 and 3) the full one-loop chargino and neutralino effects 
and (2 and 4) only the effects from $f_4$ and $f_6$. 
Similarly, in Fig.~9(c) (9(d)), the real (imaginary) part of the 
deviation in $M^{+0}$ and $M^{0-}$ is shown 
for (1 and 3) the full one-loop chargino and neutralino effects 
and (2 and 4) only the effects from $f_4$ and $f_6$. 
We note that the pure effect of the $CP$-violation can be measured 
by the difference between $M^{\pm 0}_{\tau}$ and $M^{0\mp}_{\tau}$: 
\begin{eqnarray}
M^{\pm 0}_{\tau} - M^{0 \mp}_{\tau} = - i \sqrt{2} s \gamma \tau 
\left[   \beta  F_{4,\tau} \mp F_{6,\tau} \right].  \label{mcp} 
\end{eqnarray}
As shown in Figs.~9(a) and 9(b) (9(c) and 9(d)), 
the $CP$-violating effect $M_{\rm MSSM}^{0+}-M_{\rm MSSM}^{-0}$
($M_{\rm MSSM}^{+0}-M_{\rm MSSM}^{0-}$) 
can be of the order of 0.1\% (a few times 0.1\%) 
as compared to the size of $|M_{\rm SM}^{0+}(\tau=-1)|$ 
($|M_{\rm SM}^{+0}(\tau=-1)|$)   
just after the threshold of the lightest chargino-pair production. 
The correction to $M^{+0}$ (or $M^{0-}$) is larger than that to 
$M^{0+}$ (or $M^{-0}$), 
because the SM value for the former is smaller than for the latter. 

\section{Discussion and Conclusion}
\label{sec-conclusions}
\cleqn

\hspace*{12pt}
In this paper, we have studied one-loop contributions of charginos 
and neutralinos to the helicity amplitudes of $\eeww$ in the MSSM.

The form factors are calculated 
at one loop in the $\overline{\rm MS}$ scheme.
In order to establish the validity of our one-loop calculation, 
we tested the one-loop form factors by using the BRS sum rules 
among the form factors between $\eeww$ and $e^+e^- \to \omega^+ W^-$. 
Furthermore, overall factors such as the wave-function 
renormalization factor, which cannot be tested by 
the BRS sum rules, are tested by the use of 
the decoupling property of the SUSY particles 
in the large soft-breaking mass limit. 
As pointed out in Ref.~\cite{eeww_sf}, this procedure 
for testing the one-loop calculation works well 
when we reexpand the one-loop expression of the form factors 
by the $\overline{\rm MS}$ couplings of the SM and truncate 
the higher order terms.  These tests at the numerical level 
ensure the consistency of our one-loop calculation scheme 
and our numerical program.    

The use of the SM $\overline{\rm MS}$ coupling constants as 
expansion parameters for our perturbation calculation is 
valid around and below the thresholds of the light SUSY particle 
pair production. However we have adopted this calculation 
scheme for even higher energy scales, where 
the original $\overline{\rm MS}$ scheme with the MSSM coupling 
constants should be more appropriate for the resummation of 
logarithmic terms of the type $\ln s/m_{\rm SUSY}^2$. 
In Ref.~\cite{eeww_sf}, we evaluated the error of our 
calculational scheme at high energies in the case of 
sfermion loop contributions. 
The numerical difference in $M^{00}_{\tau=-1}$ 
between our scheme and the usual $\overline{\rm MS}$ scheme 
is at most around 0.01\% for energies below a few TeV.  

We have not included the one-loop diagrams 
for the SM particles in our calculation. 
We have shown most of our results as a deviation 
from the SM prediction.

For numerical evaluation of the helicity amplitudes, 
the SUSY parameters in the chargino and neutralino sectors 
are chosen so as to satisfy the constraints from the current 
experimental data; i.e., results from electroweak 
precision measurements at the Tevatron and LEP2, direct search 
results for the chargino and neutralino at LEP2, as well as 
the current EDM data. 
Under these constraints, we took the mass of the lightest 
chargino as light as possible to obtain large corrections.  

In the $CP$ conserving case, we showed results for the 
chargino and neutralino contributions to the helicity amplitudes   
$M^{+-}_\tau$,$M^{00}_\tau$, and $M^{0+}_\tau$. 
Like the sfermion loop effect, 
the amplitude for the mode of the longitudinally polarized $W$-boson 
pair production $M_{\tau=-1}^{00}$ is found to be the most useful to study 
the chargino and neutralino contributions, having relatively large loop 
effects as compared to those for other helicity sets. 
Unlike the sfermion loop effects given in Ref.~\cite{eeww_sf},  
the enhancement at each threshold of the chargino- or 
neutralino-pair production is sharp because of the 
$s$-wave nature of the fermion-pair production threshold.
The corrections to the SM prediction for the helicity 
amplitude $M_{\tau=-1}^{00}$ can be as large as 
$-$$0.7$\% at the threshold 
of the lightest chargino-pair production for large scattering angles. 
Therefore, we found that the typical value of the chargino and neutralino 
contribution is larger than that of the sfermion contribution. 

We also studied the effects of $CP$-violating phases in the chargino 
and neutralino sectors on the helicity amplitudes. The $CP$-violating 
factors $f_4^Z$ and $f_6^V$ ($V=\gamma$ and $Z$) of the $VW^+W^-$ vertices 
are induced at one-loop level due to the triangle diagrams of 
charginos and neutralinos. Another $CP$-violating factors $f_7^V$ are not 
induced from these diagrams and remains zero.  
The size of the loop-induced form factors $f_4^{Z(1)}$ and $f_6^{V(1)}$
can be of the order of $10^{-3}$ when the $CP$-violating phases 
of the chargino and neutralino sectors are around 
$\varphi_1 \simeq \varphi_\mu \simeq 
2\pi/3$ and $4\pi/3$ for $|\mu|=130$ GeV.  
These loop-induced $CP$-violating form factors $f_4^{(1)Z}$ 
and $f_6^{(1)V}$ can affect the helicity amplitudes 
$M_{\tau=-1}^{0\pm}$ and $M_{\tau=-1}^{\mp 0}$. 
In paticular, for a large scattering angle ($\cos\theta\simeq 0$), 
the difference $M_{\tau=-1}^{0\pm}-M_{\tau=-1}^{\mp 0}$ 
measures the pure $CP$-violating effect from $f_4^{Z(1)}$ and $f_6^{V(1)}$. 
We find that the $CP$-violating effect on 
$M_{\tau=-1}^{0+}-M_{\tau=-1}^{-0}$ 
($M_{\tau=-1}^{0-}-M_{\tau=-1}^{+0}$ ) 
in the chargino and neutralino sectors 
can be as large as a few times 0.1\%  
of the SM prediction for $M_{\tau=-1}^{0+}$ ($M_{\tau=-1}^{+0}$).
 
In conclusion, the correction from the chargino and neutralino 
contributions to $\eeww$ can be as large as ${\cal O}(1\%)$ in 
amplitude, which is much larger than that of the sfermion contribution. 
The loop-induced $CP$-violating effects from the phases in the 
chargino and neutralino sectors can provide corrections of ${\cal O}(0.1\%)$ 
in amplitude.

\section*{Acknowledgments}

\hspace*{12pt}
Y.U. acknowledges the support of BMBF, contract No. 05 HT1PAA 4.
M.K. was supported by DFG under Contract No. KL 1266/1-3.


\appendix
\section{The Lagrangian}
\label{app-lagrangian}
\cleqn

\hspace*{12pt}
In this paper we are concerned with the chargino and neutralino
contributions to one-loop $\eeww$ amplitudes.  The purpose 
of this appendix is to provide all  masses, mixing angles, and 
couplings that are required to reproduce and use our results.  We begin
by discussing the chargino and neutralino mass matrices.  
We will consider two $CP$-violating phases of the $\mu$ parameter 
and the gaugino mass $M_1$, which are denoted $\varphi_\mu$ and
$\varphi_1$, respectively. 

\subsection{Chargino mass eigenstates}

\hspace*{12pt}
The chargino mass term is given by
\begin{equation}
-{\cal L} = \left( \overline{\wt{W}_{R}^{-}}\;\;\;
	\overline{\wt{H}_{uR}^{-}}  \right) M_C
	\left(
	\begin{array}{c}
	\wt{W}_{L}^{-}\\ \\ \wt{H}_{dL}^{-}
	\end{array} \right)
	+ \hc\;,
\label{equ:lag_cha_a}
\end{equation}
where the mass matrix is defined by
\begin{equation}
	M_C = \left(
	\begin{array}{cc}
	M_2 & \sqrt{2}m^{}_W c^{}_{\beta} \\
	\sqrt{2} m^{}_W s^{}_{\beta} & \mu e^{i \varphi_\mu} 
	\end{array}
	\right)\;.
\end{equation}

The matrix $M_C$ can be diagonalized by using two unitary matrices

\begin{equation}
U^{C\dagger}_R M_C U^C_L = {\rm diag}
(m^{}_{\wt{\chi}_1^-},m^{}_{\wt{\chi}_2^-})\;,
\end{equation}
where the chargino mass $m^{}_{\wt{\chi}_i^-}$ is real and positive
and has the relation $m^{}_{\wt{\chi}_1^-}<m^{}_{\wt{\chi}_2^-}$.

The mass eigenstates are defined by
\begin{equation}
\wt{\chi}_i^- = \wt{\chi}_{iL}^- + \wt{\chi}_{iR}^-\;,
\end{equation}
where 
\begin{equation}
\left( 	\begin{array}{c}
	\wt{W}_{L}^{-}\\ \wt{H}_{dL}^{-}
	\end{array} \right) = U^C_L
\left(  \begin{array}{c}
	\wt{\chi}_{1L}^{-}\\ \wt{\chi}_{2L}^{-}
	\end{array} \right),\;\;\;\;\;\;\;
\left( 	\begin{array}{c}
	\wt{W}_{R}^{-}\\ \wt{H}_{uR}^{-}
	\end{array} \right) = U^C_R
\left(  \begin{array}{c}
	\wt{\chi}_{1R}^{-}\\ \wt{\chi}_{2R}^{-}
	\end{array} \right)\;.
\label{equ:cha_rot}
\end{equation}
The mass term (\ref{equ:lag_cha_a}) is now expressed as
\begin{eqnarray}
\!\!\!\!\!\!\!\!
-{\cal L} = \left( \overline{\wt{\chi}_{1R}^{-}}\;\;\;
	\overline{\wt{\chi}_{2R}^{-}}  \right) 
        U^{C\dagger}_R M_C U^{C}_L \left(
	\begin{array}{c}
	\wt{\chi}_{1L}^{-}\\  \wt{\chi}_{2L}^{-}
	\end{array} \right)
	+ \hc 
=     \sum^2_{i=1} m^{}_{\wt{\chi}_i^-} \left(
        \overline{\wt{\chi}_{iR}^{-}}\wt{\chi}_{iL}^{-}
        + \hc \right) 
=     \sum^2_{i=1} m^{}_{\wt{\chi}_i^-} 
        \overline{\wt{\chi}_{i}^{-}}\wt{\chi}_{i}^{-}.
\label{equ:lag_cha_b}
\end{eqnarray}

\subsection{Neutralino mass eigenstates}

\hspace*{12pt}
The neutralino mass term is given by
\begin{equation}
-{\cal L} = \frac{1}{2}\left( \overline{\wt{B^{}}_{R}}\;\;\;
	\overline{\wt{W}_{R}^{3}}  \;\;\;
        \overline{\wt{H}_{dR}^{0}} \;\;\;
        \overline{\wt{H}_{uR}^{0}} 
        \right) M_N \left(
	\begin{array}{c}
	\wt{B^{}}_{L}\\ \wt{W}_{L}^{3} \\
	\wt{H}_{dL}^{0}\\ \wt{H}_{uL}^{0} 
	\end{array} \right)
	+ \hc\;,
\end{equation}

where the mass matrix is defined by

\begin{equation}
	M_C = \left(
	\begin{array}{cccc}
	M_1 e^{i \varphi_1}& 0 & -m^{}_Z s^{}_W c^{}_{\beta} & 
                                 +m^{}_Z s^{}_W s^{}_{\beta} \\
	0 & M_2 & +m^{}_Z c^{}_W c^{}_{\beta} & 
                 -m^{}_Z c^{}_W s^{}_{\beta} \\
	-m^{}_Z s^{}_W c^{}_{\beta} & 
        +m^{}_Z c^{}_W c^{}_{\beta} & 0 & -\mu e^{i \varphi_\mu} \\
	+m^{}_Z s^{}_W s^{}_{\beta} & 
        -m^{}_Z c^{}_W s^{}_{\beta} & -\mu e^{i \varphi_\mu} & 0 \\
	\end{array}
	\right).
\end{equation}

The matrix $M_N$ can be diagonalized by using two unitary matrices

\begin{equation}
U^{N\dagger}_R M_C U^N_L = {\rm diag}
(m^{}_{\wt{\chi}_1^0},m^{}_{\wt{\chi}_2^0},
 m^{}_{\wt{\chi}_3^0},m^{}_{\wt{\chi}_4^0})\,.
\end{equation}

Because the neutralinos are Majorana fermions, the mass 
matrix $M_N$ is symmetric ($M_N^T = M_N$). Therefore 
the two unitary matrices $U_L^N$ and $U_R^N$ can be chosen the same,
except for the phase matrix which makes the neutralino
mass real and positive:
\begin{eqnarray}
U^N_L = U_NP_N^*\;,  
U^N_R = U_N^*P_N\;,  
P_N = diag(e^{i \xi_1/2},e^{i \xi_2/2},e^{i \xi_3/2},e^{i \xi_4/2},)\;,
\end{eqnarray}
where $P_N$ is the phase matrix. The mass eigenstates are given by
\begin{equation}
\wt{\chi}_i^0 = \wt{\chi}_{iL}^0 + \wt{\chi}_{iR}^0 \;.
\end{equation}
The current eigenstates

\begin{equation}
X_{Li} = 
\left(  \wt{B^{}}_{L},
	\wt{W}_{L}^{3},
        \wt{H}_{dL}^{0},
        \wt{H}_{uL}^{0}
        \right) , \;\;\;\;
X_{Ri} = 
\left(   \wt{B^{}}_{R},
	 \wt{W}_{R}^{3},
         \wt{H}_{dR}^{0},
         \wt{H}_{uR}^{0}
        \right) \;,
\end{equation}
are now expressed in terms of the mass eigenstates 
$\wt{\chi}^0_{iL}$ and $\wt{\chi}^0_{iR}$, respectively, by
\begin{equation}
X_{Li} = (U^N_L)_{ij}\wt{\chi}_{jL}^0,\;\;\;\;
X_{Ri} = (U^N_R)_{ij}\wt{\chi}_{jR}^0\,.
\end{equation}

It is worth noting here that with the above phase convention 
the mass-eigenstate neutralino fields satisfy the Majorana condition
\begin{equation}
(\wt{\chi}_{iL}^0)^c = \wt{\chi}_{iR}^0\,,
\end{equation}
and hence for the four-component Majorana fields
\begin{equation}
(\wt{\chi}_{i}^0)^c = \wt{\chi}_{i}^0\,.
\end{equation}

\subsection{Chargino$-$gauge boson and neutralino$-$gauge boson interaction}
\label{subsec-masses}

\hspace*{12pt}
The interactions of gauge bosons with charginos and neutralinos 
are given by
\begin{equation}
	{\cal L_{V\wt{\chi} \wt{\chi}}} = 
	g^{\wt{\chi}^{}_1\wt{\chi}^{}_2V}_{\alpha}
	\overline{\wt{\chi}^{}_1}\gamma^{\mu}
	P_{\alpha}\wt{\chi}^{}_2 V_{\mu}\,,
\label{eq:lag_ffv}
\end{equation}
where $\chi=\chi^0$ and $\chi^-$ and $V_{\mu}= \gamma_{\mu}$ 
and $Z_{\mu}$ are implied. The couplings of the chargino-neutralino-gauge 
boson interaction are given by
\begin{subequations}
\begin{eqnarray}
	  g^{\wt{\chi}^{0}_i\wt{\chi}^{-}_j W}_{L} &=&
	 (g^{\wt{\chi}^{-}_j\wt{\chi}^{0}_i W}_{L})^* 
	=-g\left\{ (U^N_L)^*_{2i} (U^C_L)_{1j}
	+\frac{1}{\sqrt{2}} (U^N_L)^*_{3i} (U^C_L)_{2j}
	\right\}, \\ 
	  g^{\wt{\chi}^{0}_i\wt{\chi}^{-}_j W}_{R} &=&
	 (g^{\wt{\chi}^{-}_j\wt{\chi}^{0}_i W}_{R})^* 
	=-g\left\{ (U^N_R)^*_{2i} (U^C_R)_{1j}
	-\frac{1}{\sqrt{2}} (U^N_R)^*_{4i} (U^C_R)_{2j}
	\right\}.
\end{eqnarray}
\end{subequations}

The couplings of the chargino-chargino-gauge boson interaction are given by
\begin{subequations}
\begin{eqnarray}
	g^{\wt{\chi}^{-}_i\wt{\chi}^{-}_j Z}_{L} &=&
	g^{}_Z\left\{ (U^C_L)^*_{1i} (U^C_L)_{1j}
	+\frac{1}{2} (U^C_L)^*_{2i} (U^C_L)_{2j} -s^2_W\delta_{ij}
	\right\}, \\
	g^{\wt{\chi}^{-}_i\wt{\chi}^{-}_j Z}_{R} &=&
	g^{}_Z\left\{ (U^C_R)^*_{1i} (U^C_R)_{1j}
	+\frac{1}{2} (U^C_R)^*_{2i} (U^C_R)_{2j} -s^2_W\delta_{ij}
	\right\}, \\
	g^{\wt{\chi}^{-}_i\wt{\chi}^{-}_i A}_{L} &=&
	g^{\wt{\chi}^{-}_i\wt{\chi}^{-}_i A}_{R} = e.
\end{eqnarray}
\end{subequations}

The couplings of the neutralino-neutralino-gauge 
boson interaction are given by
\begin{eqnarray}
	g^{\wt{\chi}^{0}_i\wt{\chi}^{0}_j Z}_{L} =
	-\frac{1}{2} g^{}_Z\left\{ (U^N_L)^*_{3i} (U^N_L)_{3j}
	-(U^N_L)^*_{4i} (U^N_L)_{4j} 
	\right\}, \;\;  
	g^{\wt{\chi}^{0}_i\wt{\chi}^{0}_j Z}_{R} =
	-g^{\wt{\chi}^{0}_j\wt{\chi}^{0}_i Z}_{L}.
\end{eqnarray}

The interaction with charge conjugated fermions of 
Eq.~(\ref{eq:lag_ffv}) can be rewritten by
\begin{equation}
	{\cal L} = 
	g^{\wt{\chi}^{c}_1\wt{\chi}^{c}_2V}_{\alpha}
	\overline{\wt{\chi}^{c}_1}\gamma^{\mu}
	P_{\alpha}\wt{\chi}^{c}_2 V_{\mu},
\label{eq:lag_fcfcv}
\end{equation}
where the coupling $g^{\wt{\chi}^{c}_1\wt{\chi}^{c}_2V}_{\alpha}$
is related to the coupling $g^{\wt{\chi}^{}_1\wt{\chi}^{}_2V}_{\alpha}$
by
\begin{eqnarray}
g^{\wt{\chi}^{c}_1\wt{\chi}^{c}_2V}_{L} = -
g^{\wt{\chi}^{ }_2\wt{\chi}^{ }_1V}_{R}\;, \;\;   
g^{\wt{\chi}^{c}_1\wt{\chi}^{c}_2V}_{R} = -
g^{\wt{\chi}^{ }_2\wt{\chi}^{ }_1V}_{L}\;.
\end{eqnarray}
The minus sign arises because of the charge conjugation of the
vector current.

\subsection{Chargino$-$Goldstone boson and neutralino$-$Goldstone 
boson interaction}

\hspace*{12pt}
The interactions of the Goldstone boson with charginos and neutralinos 
are given by
\begin{equation}
	{\cal L_{\omega \wt{\chi} \wt{\chi}}} = 
	g^{\wt{\chi}^{}_1\wt{\chi}^{}_2\omega}_{\alpha}
	\overline{\wt{\chi}^{}_1}
	P_{\alpha}\wt{\chi}^{}_2 \omega .
\label{eq:lag_ffs}
\end{equation}

The chargino-neutralino-Goldstone boson couplings are given by
\begin{subequations}
\begin{eqnarray}
	 g^{\wt{\chi}^{0}_i\wt{\chi}^{-}_j \omega^-}_{L} \!\!\!\!=
	(g^{\wt{\chi}^{-}_j\wt{\chi}^{0}_i \omega^-}_{R})^* &\!\!=\!&
	 -i\frac{g}{\sqrt{2}} \left\{
                 (U^N_R)_{2i}^* (U^C_L)_{2j}
        -\sqrt{2}(U^N_R)_{3i}^* (U^C_L)_{1j}
        +t^{}_W  (U^N_R)_{1i}^* (U^C_L)_{2j}
	\right\} \cos\beta\,, \\
	 g^{\wt{\chi}^{0}_i\wt{\chi}^{-}_j \omega^-}_{R} \!\!\!\!=
	(g^{\wt{\chi}^{-}_j\wt{\chi}^{0}_i \omega^-}_{L})^* &\!\!=\!&
	 -i\frac{g}{\sqrt{2}} \left\{
                 (U^N_L)_{2i}^* (U^C_R)_{2j}
        +\sqrt{2}(U^N_L)_{4i}^* (U^C_R)_{1j}
        +t^{}_W  (U^N_L)_{1i}^* (U^C_R)_{2j} 
	\right\} \sin\beta\,. \makebox[10mm]{}
\end{eqnarray}
\end{subequations}

The interaction with charge conjugated fermions of 
Eq.~(\ref{eq:lag_ffs}) can be rewritten as
\begin{equation}
	{\cal L} = 
	g^{\wt{\chi}^{c}_1\wt{\chi}^{c}_2\omega}_{\alpha}
	\overline{\wt{\chi}^{c}_1}
	P_{\alpha}\wt{\chi}^{c}_2 \omega \;,
\label{eq:lag_fcfcs}
\end{equation}
where the coupling $g^{\wt{\chi}^{c}_1\wt{\chi}^{c}_2\omega}_{\alpha}$
is related to the coupling 
$g^{\wt{\chi}^{}_1\wt{\chi}^{}_2\omega}_{\alpha}$ by
\begin{eqnarray}
g^{\wt{\chi}^{c}_1\wt{\chi}^{c}_2V}_{L} = 
g^{\wt{\chi}^{ }_2\wt{\chi}^{ }_1V}_{L}\;, \;\;  
g^{\wt{\chi}^{c}_1\wt{\chi}^{c}_2V}_{R} = 
g^{\wt{\chi}^{ }_2\wt{\chi}^{ }_1V}_{R}\;.
\end{eqnarray}

\cleqn
\section{Chargino and neutralino effects on the form factors}

\subsection{Form factors $\widetilde{F}_{i,\tau}$ and 
$F_{i,\tau}^{\rm (1)ext}$}

\hspace*{12pt}
The $\widetilde{F}_{i,\tau}$ are expressed by  
\begin{eqnarray}
\nonumber
&&\makebox[-1cm]{} 
\widetilde{F}_{i,\tau}(s,t)  =  \frac{\hatesq}{s}\Bigg\{\bigg[Q_e \Big(
        1 - \pitggg(s) + \Gamma_1^{\,e}(s)\Big) + T^3_e\, 
        \overline{\Gamma}\mbox{}^{\,e}_2(s) 
        \bigg]f_i^{\gamma\,(0)} 
        +  Q_e f_i^{\gamma\,(1)}(s) \Bigg\}
\\ \nonumber && 
\makebox[-0.5cm]{} 
        + \frac{\hatgsq}{s - (m_W^2/\hat{c}^2)}
        \Bigg\{\bigg[\big(T^3_e-\hatssq Q_e\big)
        \Big( 1 + 
             \frac{\Delta}{s - m_W^2/\hat{c}^2}
 - \pitzzz(s) + \Gamma_1^{\,e}(s) \Big) 
        + T^3_e \Big( \hat{c}^2 \overline{\Gamma}_2^{\,e}(s)  
+   \Gamma_3^{\,e}(s) \Big) \\ \nonumber&& 
\makebox[1.6cm]{}  
+ \Gamma_4^{\,e}(s) \bigg]f_i^{Z\,(0)} 
        + \big(T^3_e-\hatssq Q_e\big) f_i^{Z\,(1)}(s)
        - \frac{\hats}{\hatc} \bigg[ Q_e \hatcsq f_i^{Z\,(0)} 
        + \big(T^3_e-\hatssq Q_e\big) f_i^{\gamma\,(0)} \bigg] \pitgzg(s)
           \Bigg\}
\\ && 
\makebox[-0.5cm]{} 
        + \frac{T^3_e\hatgsq}{2t} 
\Big[ f_i^{t\,(0)} + \Gamma^{e\nu W}(t) + \overline{\Gamma}^{e\nu W}(t) \Big] 
        + F_{i,\tau}^{[\rm Box]}(s,t)\;,
\label{big-F}
\end{eqnarray}
where $i = 1$ - $16$. The two-point functions $\pitggg$, $\pitgzg$,
and $\pitzzz$ are given in Appendix B~3.

The vertex coefficients $f_i^{V}$ are  
divided into the tree contribution and the one-loop vertex contribution 
according to Eq.~(\ref{ff01}),   
\begin{equation}
 f_i^V(s) = f_i^{V\,(0)} + f_i^{V\,(1)}(s)\;,
\end{equation}
where $V = \gamma$ and $Z$.  
The nonzero tree-level values $f_i^{V\,(0)}$ are given in 
Table~\ref{table-smf0}.  
\begin{table}[t]
\begin{center}
\begin{tabular}{|c||cccccccccccccccc|}\hline
i & 1 & 2 & 3 & 4 & 5 & 6 & 7 & 8 & 9 & 10 & 11 & 12 & 13 & 14 & 15 & 16 \\ 
\hline \hline
$f_{i}^{\gamma\,(0)}$ &1&&2&& &&& &&$-1$&&&1&&& \\ 
$f_{i}^{Z\,(0)}$      &1&&2&& &&& &&$-1$&&&1&&& \\ 
$f_{i}^{t\,(0)}$      &1&&2&&1&&&1&&$-2$&&&2&&& \\ \hline
\end{tabular}
\end{center}
\caption{Explicit values for 
$f_i^{\gamma\,(0)}$, $f_i^{Z\,(0)}$, and $f_i^{t\,(0)}$ in 
Eq.~\eq{big-F}.  
Only nonzero values are shown.}
\label{table-smf0} 
\end{table}
The vertex functions for the $Vee$ vertex, denoted by 
$\Gamma_1^{\,e}$, $\overline{\Gamma}\mbox{}^{\,e}_2$, $\Gamma_3^{\,e}$, and 
$\Gamma_4^{\,e}$, also appear in the 
$e^-e^+\rightarrow f\overline{f}$ amplitudes~\cite{hhkm94}. 
The vertex functions 
$\Gamma^{\,e\nu W}$ and $\overline{\Gamma}\mbox{}^{\,e\nu W}$ 
appear in charged current processes; they contain $\nu e W$ vertex corrections 
as well as two-point function corrections 
for the external electrons and $W$ bosons and the internal neutrino 
propagator. 
Finally, the 
$F_{i,\tau}^{[\rm Box]}$ terms account for contributions of box diagrams.
In the limit of heavy SUSY particles except for the 
chargino and neutralino that 
we study in this paper, all these vertex functions $\Gamma_1^{\,e}$, 
$\overline{\Gamma}\mbox{}^{\,e}_2$, $\Gamma_3^{\,e}$,
$\Gamma_4^{\,e}$,$\Gamma^{\,e\nu W}$, and 
$\overline{\Gamma}\mbox{}^{\,e\nu W}$ 
and the box corrections are 
small and we can set them to zero.

Next, for the part of the corrections to external $W$-boson lines,  
$F_{i,\tau}^{\rm (1)ext}$, we have only to discuss the cases in which   
all the external $W$ bosons are physical ($\lambda$ or $\overline{\lambda} = 
0, \pm 1$);  
\begin{eqnarray}
&&\makebox[-1cm]{}  
F_{i,\tau}^{\rm (1)ext}(s,t)  = \Bigg[  \frac{\hatesq}{s} 
              Q_e f_i^{\gamma\,(0)} 
        + \frac{\hatgsq}{s - (m_W^2/\hat{c}^2)}
        \big(T^3_e-\hatssq Q_e\big)
                       f_i^{Z\,(0)} 
        + \frac{T^3_e\hatgsq}{2t} 
 f_i^{t\,(0)}  \Bigg] 
  \delta Z_W , 
\label{big-Fext}
\end{eqnarray}
where $i = 1$ - $9$ and 
$\delta Z_W$ is the wave-function renormalization factor of physical 
$W$ bosons with helicities $\lambda$ or $\overline{\lambda} = 0, \pm$, 
and its chargino and neutralino one-loop contributions are 
given in Appendix B~3.

\subsection{Form factors $H_{i,\tau}(s,t)$}
\label{subsec-ff-eewx}

\hspace*{12pt}
The $H_{i,\tau}(s,t)$ are expressed by
\begin{eqnarray}
\nonumber
  \lefteqn{ H_{i,\tau}(s,t) = 
\frac{\hatesq}{s}\Bigg\{\bigg[ Q_e \Big( 1
- \pitggg(s) + \Gamma_1^{\,e}(s)
\Big) + T^3_e \overline{\Gamma}\mbox{}^{\,e}_2(s) \bigg]
h_i^{\gamma\,(0)} 
+ Q_e h_i^{\gamma\,(1)}(s) \Bigg\}}
&&\\
\nonumber && \makebox[-1.5cm]{} 
+ \frac{\hatgsq}{s  - \mwsq/\hat{c}^2}
\Bigg\{ \bigg[ (T^3_e-\hatssq Q_e)\Big( 
 1  + 
\frac{\Delta}{s-m_W^2/\hat{c}^2}
  - \pitzzz(s) + \Gamma_1^{\,e}(s)
\Big) 
+ T^3_e \Big( \hat{c}^2 \overline{\Gamma}_2^{\,e}(s) + \Gamma_3^{\,e}(s) \Big)
\nonumber \\&&
\makebox[-1.5cm]{}
+ \Gamma_4^{\,e}(s)\bigg] h_i^{Z\,(0)}  + 
(T^3_e-\hatssq Q_e) h_i^{Z\,(1)}(s)  
- \frac{\hats}{\hatc} 
\bigg[ (T^3_e-\hatssq Q_e)h_i^{\gamma\,(0)} 
+ Q_e \hat{c}^2 h_i^{Z\,(0)} \bigg]
\pitgzg(s) \Bigg\} \nonumber \\
&&+ \frac{\hatgsq T_e^3}{2 t} \Gamma^{e\nu\omega}
+ H_{i,\tau}^{\,\boxes}(s,t) \;.
\label{big-H}
\end{eqnarray}

The vertex form factors are written as 
the sum of the tree-level and one-loop contributions by
\begin{equation}
h_i^V(s) = h_i^{V\,(0)} + h_i^{V\,(1)}(s)\;,
\end{equation}
for $V = \gamma$, $Z$. 
The tree-level form-factor coefficients ${h}_i^{\,V\,(0)}$ are given by 
$h_i^{\,\gamma\,(0)} =  \delta_{i\; 1}$ and 
$h_i^{\, Z ,(0)} =  -({\hatssq}/{\hatcsq}) \delta_{i\; 1}$. 
The $h_i^{V\,(1)}(s)$ come from the one-loop 
1PI $VW\chi$ vertex corrections. 
The chargino and neutralino contributions to $h_i^{V\,(1)}(s)$ 
are shown in Appendix B~5 \,. 
All the one-loop vertex contributions $\Gamma^e_1$, 
$\overline{\Gamma}^e_2$, $\Gamma^e_3$, $\Gamma^e_4$, and 
$\Gamma^{e\nu\omega}$ and the box diagrams $H_{i,\tau}^{\rm [Box]}$
that connect with initial $e^\pm$ lines  
turn out to be zero for the chargino and neutralino contributions.

\subsection{Two-point functions}

\hspace*{12pt}
The explicit forms of the two-point functions of $\Pi_T^{\gamma\gamma}$, 
$\Pi_T^{\gamma Z}$, $\Pi_T^{ZZ}$, and $\Pi_T^{WW}$ are the 
following~\cite{dobado}.

\bsub
\bea
\Pi_T^{\gamma\gamma} &=& \frac{\ehat^2}{16\pi^2}
 8q^2B_3 (q^2, \mch{i}, \mch{i})\;, 
\\
\Pi_T^{Z\gamma} &=& \frac{\ehat\gzhat}{16\pi^2}
	 \biggl\{
	  (D_L)_{ii} 	
	+ (D_R)_{ii} 
	-2\shat^2\biggr\}
	4 q^2 B_3 (q^2 , \mch{i}, \mch{i})\;, 
\\
\Pi_T^{ZZ} &=& \frac{\gzhatsq}{16\pi^2} \biggl[
	\biggl\{ -\shat^2 | (D_L)_{ii} |^2
	          -\shat^2 | (D_R)_{ii} |^2
	          +\shat^4 \biggr\}8q^2B_3(q^2, \mch{i}, \mch{i}) 
\nonumber\\ &&
	2\biggl\{ |(D_L)_{ij}|^2 + |(D_R)_{ij}|^2 \biggr\}
	(2q^2 B_3 - B_4) (q^2, \mch{i}, \mch{j}) 
\nonumber\\ 
	&&
	+ 2 \mch{i} \mch{j} 
	\biggl\{ (D_L)_{ij} (D_R)_{ij}^*  
	+ (D_L)_{ij}^* (D_R)_{ij} \biggr\}
	B_0 (q^2, \mch{i}, \mch{j}) 
\nonumber\\ 
	&&
	+ \biggl\{ |(N_L)_{ij}|^2 + |(N_R)_{ij}|^2\biggr\}
	(2q^2 B_3 - B_4)(q^2,\mn{i}, \mn{j})
\nonumber\\ 
	&&
	+ \mn{i} \mn{j} \biggl\{ (N_L)_{ij} (N_R)_{ij}^*
		+ (N_L)_{ij}^* (N_R)_{ij} \biggr\}
	B_0 (q^2,\mn{i}, \mn{j})\biggr]\;, 
\\
\Pi_T^{WW} &=& \frac{\hat{g}^2}{16\pi^2}  
	2 \biggl[ \biggl\{ |(C_{L})_{ij}|^2 + |(C_{R})_{ij}|^2\biggr\}
	(2q^2 B_3 - B_4)(q^2, \mn{i}, \mch{j}) 
\nonumber\\
	&& 
	+ \mn{i} \mch{j} 
 	\biggl\{ (C_{L})_{ij} (C_{R})_{ij}^* 
	+ (C_{L})_{ij}^* (C_{R})_{ij} \biggr\}
	B_0(q^2, \mn{i}, \mch{j}) 
	\biggr]\;, 
\eea
\esub
where 
\bsub
\bea
(C_{\alpha})_{ij} 
	&=& (\nunitary{\alpha})^*_{2i} (\cunitary{\alpha})_{1j}
 + \frac{1}{\sqrt{2}} (\nunitary{\alpha})^*_{3i}(\cunitary{\alpha})_{2j},
\label{eq:ino_coupling_one}
\\
(D_{\alpha})_{ij} 
	&=& (\cunitary{\alpha})^*_{1i} (\cunitary{\alpha})_{1j} 
	+ \half (\cunitary{\alpha})^*_{2i} (\cunitary{\alpha})_{2j},  
\label{eq:ino_coupling_two}
\\
(N_{L})_{ij} &=& -(N_{R})^*_{ij} = 
	\half \biggl\{ (\nunitary{L})^*_{3i}(\nunitary{L})_{3j} - 
	(\nunitary{L})^*_{4i}(\nunitary{L})_{4j} \biggr\}.
\eea
\esub
Here $\alpha=L$ or $R$ in Eqs. (\ref{eq:ino_coupling_one}) 
and (\ref{eq:ino_coupling_two}). 

The two-point functions
$\Pi^{QQ}_{T}$, $\Pi^{3Q}_{T}$, $\Pi^{33}_{T}$, and
$\Pi^{11}_{T}$ can be written as
$\Pi^{\gamma\gamma}_{T}$, 
$\Pi^{\gamma Z}_{T}$,
$\Pi^{Z Z}_{T}$, and $\Pi^{W W}_{T}$.
The relationships among them 
are shown in Eqs.~(A1) of Ref.~\cite{hhkm94}.
The gauge boson two-point functions $\Pi^{\gamma\gamma}_{T,\gamma}(\qsq)$, 
$\Pi^{\gamma Z}_{T,\gamma}(\qsq)$, and $\Pi^{ZZ}_{T,Z}(\qsq)$ can be 
obtained from the transverse part of the vacuum polarization tensor
\begin{equation}
\label{defn-pitv}
\Pi^{V_1V_2}_{T,V_3}(\qsq) = 
\frac{\Pi^{V_1V_2}_T(\qsq) - \Pi^{V_1V_2}_T(m^2_{V_3})}  {\qsq - m^2_{V_3}}\;,
\end{equation}
where $V_1$, $V_2$, and $V_3$ denote the gauge bosons.
The one-loop chargino and neutralino contributions to the 
wave-function renormalization factor of the physical $W$
boson are obtained from the $WW$ two-point function $\pitww$
\begin{equation}
Z_{W}^\frac{1}{2} = 1 - \frac{1}{2} 
       \left.  \frac{d}{d q^2} \; \pitww(q^2) \right|_{q^2=\mwsq}\;, \;\;
\;\;
\delta Z_{W}^\frac{1}{2} = Z_{W}^\frac{1}{2}  - 1 \; .
\end{equation}

\subsection{$VW^+W^-$ triangle vertex functions}

\begin{figure}[t]
\begin{center}
\includegraphics*[width=9cm]{10.epsi} 
\caption{ Mass and momentum assignments for the calculation of the 
chargino and neutralino triangle graph are shown. The arrows in 
the $W$ lines indicate the flow of a negative electric charge.}
\label{fig:feyn_a}
$\begin{array}{lr}
\includegraphics*[width=3.5cm]{11a.epsi}
\;\;\;\; & \;\;\;\;
\includegraphics*[width=3.5cm]{11b.epsi} 
\end{array}$
\end{center}
\caption{Feynman graphs contributing to the $VW^+W^-$ vertex are shown.
The mass and momentum assignments are shown in 
Fig.~\ref{fig:feyn_a}.
}
\label{fig:feyn_b}
\vspace*{1cm} 
\end{figure}

\hspace*{12pt}
We discuss one-loop chargino and neutralino contributions to the
$VW^+W^-$ triangle vertex diagram~\cite{dobado2}. The assignments of mass, 
momentum, and helicity of the couplings are shown
in Fig.~\ref{fig:feyn_a}. 

For evaluation of the loop integrals we use the convention of 
incoming momenta; hence we use $p_1 = -p$ and $p_2 = -\overline{p}$ where
$p$ and $\overline{p}$ are defined in Fig.~\ref{fig:feyn_a}.
Dropping the coupling factors, the tensor structure of the 
triangle diagram (Fig.~11) is given by 
\begin{equation}
T_{\rm INOT}^{\mu\alpha\beta}(p_1,p_2,m_1^2,m_2^2,m_3^2) = 
\sum_i C^{\rm INO}_i(p_1,p_2,m_1^2,m_2^2,m_3^2) T_i^{\mu\alpha\beta}
\end{equation}
where the tensor structures of $T_i^{\mu\alpha\beta}$ are listed in
Ref.~\cite{brs}. The subscript ``INOT''  denotes ``INO triangle''
contributions.
The  nonzero $C^{\rm INO}_i$ are given by
%
%
%
%
%
\begin{subequations}
\begin{eqnarray}
C^{\rm INO}_1 = && 
\sum_{\tau =\sigma =\rho}  m^2_W
\left( C_{31}-C_{32}-3C_{33}+3C_{34}-\frac{2-D}{m^2_W}C_{35}
+\frac{2-D}{m^2_W}C_{36}+C_{21}+C_{22}-2C_{23} \right.
\nonumber\\ && \left.
+\frac{2-D}{m^2_W}C_{24}
+\frac{s}{m^2_W}(C_{33}-C_{34}-C_{22}-C_{12}    )\right)
+\sum_{\tau =\sigma =-\rho}m^{}_im^{}_j(-C_{11}+C_{12})\nonumber\\
&&+\sum_{\tau =-\sigma =\rho}m^{}_jm^{}_k(-C_{11}+C_{12})
+\sum_{\tau =-\sigma =-\rho}m^{}_km^{}_i(C_{11}-C_{12}+C_{0})\,, \\
C^{\rm INO}_2 = &&
\sum_{\tau =\sigma =\rho} m^2_W
(4C_{33}-4C_{34}-4C_{22}+4C_{23})\,,\\
C^{\rm INO}_3 = &&
\sum_{\tau =\sigma =\rho} m^2_W
\left(-C_{31}+C_{32}+3C_{33}-3C_{34}
+\frac{2-D}{m^2_W}C_{35}-\frac{2-D}{m^2_W}C_{36} \right.
\nonumber\\ && 
-3C_{21}-3C_{22}+6C_{23}
+\frac{6-3D}{m^2_W}C_{24}-2C_{11}+2C_{12} \nonumber \\ && \left.
+\frac{s}{m^2_W}(-C_{33}+C_{34}+C_{22}-2C_{23}-C_{12})\right)
+\sum_{\tau =\sigma =-\rho}m^{}_im^{}_j(C_{11}+C_{12}+2C_{0})\nonumber\\
&&+\sum_{\tau =-\sigma =\rho}m^{}_jm^{}_k(-C_{11}-C_{12})
+\sum_{\tau =-\sigma =-\rho}m^{}_km^{}_i(C_{11}-C_{12}+C_{0})\,,\\
C^{\rm INO}_4 = &&
\sum_{\tau =\sigma =\rho} im^2_W
\left(-C_{31}-C_{32}+C_{33}
+C_{34}+\frac{2-D}{m^2_W}C_{35}+\frac{2-D}{m^2_W}C_{36}-C_{21} \right.
\nonumber \\&& \left.
-C_{22}+2C_{23}+\frac{2-D}{m^2_W}C_{24}
+\frac{s}{m^2_W}(-C_{33}-C_{34}-C_{22}-2C_{23}-C_{12})\right) 
\nonumber\\ 
&&+\sum_{\tau =\sigma =-\rho}im^{}_im^{}_j(C_{11}-C_{12})
+\sum_{\tau =-\sigma =\rho}im^{}_jm^{}_k(-C_{11}+C_{12})\nonumber\\
&&+\sum_{\tau =-\sigma =-\rho}im^{}_km^{}_i(C_{11}+C_{12}+C_{0})\,,\\
C^{\rm INO}_5 = &&
\sum_{\tau =\sigma =\rho} \tau m^2_W 
\left(C_{31}-C_{32}-3C_{33}+3C_{34}+\frac{2+D}{m^2_W}C_{35}
-\frac{2+D}{m^2_W}C_{36} \right.
\nonumber\\&& \left.
+C_{21}+C_{22}-2C_{23}
+\frac{2}{m^2_W}C_{24}
+\frac{s}{m^2_W}(C_{33}-C_{34}-C_{22}+2C_{23}+C_{12})\right)
\nonumber\\
&&+\sum_{\tau = \sigma =-\rho}\tau m^{}_im^{}_j (-C_{11}+C_{12})
+\sum_{\tau =-\sigma = \rho}\tau m^{}_jm^{}_k (-C_{11}+C_{12})\nonumber\\
&&+\sum_{\tau =-\sigma =-\rho}\tau m^{}_km^{}_i (C_{11}-C_{12}+C_{0})\,,\\
C^{\rm INO}_6 = &&
\sum_{\tau =\sigma =\rho} -i\tau m^2_W
\left(C_{31}+C_{32}-C_{33}-C_{34}+\frac{2+D}{m^2_W}C_{35}
+\frac{2+D}{m^2_W}C_{36} \right.
\nonumber \\&&
+3C_{21}-C_{22}-2C_{23}
-\frac{2-2D}{m^2_W}C_{24}+2C_{11}-2C_{12}
\nonumber \\&& \left.
+\frac{s}{m^2_W}(C_{33}+C_{34}+C_{22}+2C_{23}+C_{12})\right) 
+\sum_{\tau = \sigma =-\rho}i\tau m^{}_im^{}_j (C_{11}+C_{12}+2C_{0})
\nonumber\\
&&+\sum_{\tau =-\sigma = \rho}i\tau m^{}_jm^{}_k (C_{11}+C_{12})
+\sum_{\tau =-\sigma =-\rho}i\tau m^{}_km^{}_i (-C_{11}-C_{12}-C_{0})\,,\\
C^{\rm INO}_{10} = &&
\sum_{\tau =\sigma =\rho} m^2_W
\left(2C_{31}-2C_{32}-6C_{33}+6C_{34}-\frac{4-2D}{m^2_W}C_{35}
+\frac{4-2D}{m^2_W}C_{36} \right.
\nonumber \\ &&
+2C_{21}+2C_{22}-4C_{23}
-\frac{4-2D}{m^2_W}C_{24}
\nonumber \\ && \left.
+\frac{s}{m^2_W}(2C_{33}-2C_{34}+2C_{21}-2C_{22}+2C_{23}+2C_{11})\right)
\nonumber\\
&&+\sum_{\tau = \sigma =-\rho} m^{}_im^{}_j (2C_{11}-2C_{12})
+\sum_{\tau =-\sigma = \rho} m^{}_jm^{}_k (-2C_{11}+2C_{12})\nonumber\\
&&+\sum_{\tau =-\sigma =-\rho} m^{}_km^{}_i (-2C_{11}+2C_{12}-2C_{0})\,,\\
C^{\rm INO}_{11} = &&
\sum_{\tau =\sigma =\rho} m^2_W 
(-4C_{31}+8C_{33}-4C_{34}-6C_{21}-4C_{22}+10C_{23}
-2C_{11}+2C_{12})\,, \\
C^{\rm INO}_{12} = &&
\sum_{\tau =\sigma =\rho} \tau m^2_W 
(2C_{21}-2C_{23}+2C_{11}-2C_{12})\,,\\
C^{\rm INO}_{13} = &&
\sum_{\tau =\sigma =\rho} m^2_W
\left(-2C_{31}+2C_{32}+6C_{33}-6C_{34}+\frac{4-2D}{m^2_W}C_{35}
-\frac{4-2D}{m^2_W}C_{36} \right.
\nonumber \\ && \left.
-2C_{21}-2C_{22}+4C_{23}
+\frac{4-2D}{m^2_W}C_{24}
+\frac{s}{m^2_W}(-2C_{33}+2C_{34}-2C_{23}-2C_{12})\right) \nonumber\\
&&+\sum_{\tau = \sigma =-\rho} m^{}_im^{}_j (2C_{11}-2C_{12})
+\sum_{\tau =-\sigma = \rho} m^{}_jm^{}_k (-2C_{11}+2C_{12})\nonumber\\
&&+\sum_{\tau =-\sigma =-\rho} m^{}_km^{}_i (2C_{11}-2C_{12}+2C_{0})\,,\\
C^{\rm INO}_{14} = &&
\sum_{\tau =\sigma =\rho} m^2_W
(-4C_{32}-4C_{33}+8C_{34}+2C_{22}-2C_{23})\,,\\
C^{\rm INO}_{15} = &&
\sum_{\tau =\sigma =\rho} \tau m^2_W\
(-2C_{22}+2C_{23})\,, \\
C^{\rm INO}_{16} = &&
\sum_{\tau =\sigma =\rho} 4 m^2_W
(-C_{31}+C_{32}+3C_{33}-3C_{34}
-2C_{21}-2C_{22}+4C_{23}
-C_{11}+C_{12})\,.
\end{eqnarray}
\end{subequations}

The next step is to provide the correct couplings and masses and
then sum over all triangle graphs.  For the $\gamma W^+W^-$ vertex,
\begin{equation}
f_i^{\gamma\,(1)\,\rm INOT} = -\frac{1}{16\pi^2\hate}
g^{\wt{\chi}^-_i\wt{\chi}^-_iA}_{\tau}
g^{\wt{\chi}^-_i\wt{\chi}^0_jW}_{\sigma}
g^{\wt{\chi}^0_j\wt{\chi}^-_iW}_{\rho}
C^{\rm INO}_{i}(p_1,p_2,m^2_{\wt{\chi}^-_i},m^2_{\wt{\chi}^0_j},
m^2_{\wt{\chi}^-_i})\,,
\end{equation}
where summation over charginos and neutralinos is implied.  
For the $ZW^+W^-$ vertex,
\begin{eqnarray}
f_i^{Z\,(1)\,\rm INOT} &=&
-\frac{1}{16\pi^2\hatgz\hatcsq}\left\{
g^{\wt{\chi}^-_i\wt{\chi}^-_kZ}_{\tau}
g^{\wt{\chi}^-_k\wt{\chi}^0_jW}_{\sigma}
g^{\wt{\chi}^0_j\wt{\chi}^-_iW}_{\rho}
C^{\rm INO}_{i}(p_1,p_2,m^2_{\wt{\chi}^-_i},m^2_{\wt{\chi}^0_j},
m^2_{\wt{\chi}^-_k}) \right. \nonumber\\ 
&&\left. \;\;\;\;\;\;\; +
g^{\wt{\chi}^0_i\wt{\chi}^0_kZ}_{\tau}
g^{\wt{\chi}^0_k\wt{\chi}^+_jW}_{\sigma}
g^{\wt{\chi}^+_j\wt{\chi}^0_iW}_{\rho}
C^{\rm INO}_{i}(p_1,p_2,m^2_{\wt{\chi}^0_i},m^2_{\wt{\chi}^+_j},
m^2_{\wt{\chi}^0_k})\right\} \,,
\end{eqnarray}
where summation over charginos and neutralinos is implied.

\subsection{$V\omega^+W^-$ triangle vertex functions}

\begin{figure}[t]
\begin{center}
\includegraphics*[width=8.5cm]{12.epsi} 
\caption{ Mass and momentum assignments for the calculation of the 
chargino and neutralino triangle graph are shown. The arrows in 
the $W$ lines indicate the flow of a negative electric charge.}
\label{fig:feyn_c}
$\begin{array}{lr}
\includegraphics*[width=3.5cm]{13a.epsi}
\;\;\; & \;\;\;
\includegraphics*[width=3.5cm]{13b.epsi} 
\end{array}$
\end{center}
\caption{Feynman graphs contributing to the $V\omega^+W^-$ vertex are shown.
The mass and momentum assignments are shown in 
Fig.~\ref{fig:feyn_c}.
}
\label{fig:feyn_d}
\vspace*{1cm} 
\end{figure}

\hspace*{12pt}
We discuss one-loop chargino and neutralino contributions to the
$V\omega^+W^-$ triangle vertex diagram. The assignments of mass, 
momentum, and helicity of the couplings are in Fig.~\ref{fig:feyn_c}. 
Dropping the coupling factors, the tensor structure of the 
triangle diagram (Fig.~13) is given by 
\begin{equation}
S_{\rm INOT}^{\mu\alpha}(p_1,p_2,m_1^2,m_2^2,m_3^2) = 
\sum_i c^{\rm INO}_i(p_1,p_2,m_1^2,m_2^2,m_3^2) S_i^{\mu\alpha}\;,
\end{equation}
where the tensor structures of $S_i^{\mu\alpha}$ are listed in
Ref.~\cite{brs}. The  nonzero $S^{\rm INO}_i$ are given by
%
%
%
%
%
%
\begin{subequations}
\begin{eqnarray}
c^{\rm INO}_1 = && 
\sum_{\tau =-\sigma =-\rho}  2 m^{}_i m^{}_W
\left(C_{21}+C_{22}-2C_{23}+\frac{D}{m^2_W}C_{24}
+C_{11}-C_{12}
+\frac{s}{m^2_W}(2C_{23}+C_{11}+2C_{12}+C_{0}) \right) 
\nonumber\\
&& +\sum_{\tau =-\sigma =\rho}  m^{}_j m^{}_W
\left(-2C_{21}-2C_{22}+4C_{23}+\frac{4-2D}{m^2_W}C_{24}
+\frac{s}{m^2_W}(-2C_{23}-C_{11}-C_{12}) \right)
\nonumber\\
&& +\sum_{\tau = \sigma =\rho}  m^{}_k m^{}_W
\left(-2C_{21}-2C_{22}+4C_{23}+\frac{4-2D}{m^2_W}C_{24}
-2C_{11}+2C_{12}
+\frac{s}{m^2_W}(-2C_{23}-C_{12}) \right) \nonumber\\
&& +\sum_{\tau = \sigma =-\rho} m^{}_i m^{}_j m^{}_k 
\frac{1}{m^{}_W}2C_{0}\,, \\
c^{\rm INO}_2 = && 
\sum_{\tau =-\sigma =-\rho} m^{}_i m^{}_W (C_{11}+C_{0})
+\sum_{\tau =-\sigma =\rho}  m^{}_j m^{}_W
(2C_{22}-2C_{23}-C_{11}+C_{12})\nonumber \\
&& +\sum_{\tau = \sigma =\rho}  m^{}_k m^{}_W
(2C_{22}-2C_{23}-C_{12})\,,\\
c^{\rm INO}_3 = && 
\sum_{\tau =-\sigma =-\rho} \tau m^{}_i m^{}_W (C_{11}+C_{0}) 
+\sum_{\tau =-\sigma =\rho} \tau m^{}_j m^{}_W
(-C_{11}+C_{12})
+\sum_{\tau = \sigma =\rho} \tau m^{}_k m^{}_W
(C_{12})\,, \\
c^{\rm INO}_4 = && 
\sum_{\tau =-\sigma =-\rho} m^{}_i m^{}_W(-2C_{11}-2C_{0}) 
 +\sum_{\tau =-\sigma =\rho} \tau m^{}_j m^{}_W
(2C_{21}-2C_{22}+2C_{11}-2C_{12}) \nonumber\\
&& +\sum_{\tau = \sigma =\rho} \tau m^{}_k m^{}_W
(2C_{21}-2C_{22}+2C_{11})\,. 
\end{eqnarray}
\end{subequations}

The next step is to provide the correct couplings and masses and
then sum over all triangle graphs.  For the $\gamma \omega^+W^-$ vertex,
\begin{equation}
h_i^{\gamma\,(1)\,\rm INOT} = -\frac{1}{16\pi^2\hate}
g^{\wt{\chi}^-_i\wt{\chi}^-_iA}_{\tau}
g^{\wt{\chi}^-_i\wt{\chi}^0_j\omega}_{\sigma}
g^{\wt{\chi}^0_j\wt{\chi}^-_iW}_{\rho}
C^{\rm INO}_{i}(p_1,p_2,m^2_{\wt{\chi}^-_i},m^2_{\wt{\chi}^0_j},
m^2_{\wt{\chi}^-_i})\,,
\end{equation}
where summation over charginos and neutralinos is implied.  
For the $Z\omega^+W^-$ vertex,
\begin{eqnarray}
h_i^{Z\,(1)\,\rm INOT} &=&
-\frac{1}{16\pi^2\hatgz\hatcsq}\left\{
g^{\wt{\chi}^-_i\wt{\chi}^-_kZ}_{\tau}
g^{\wt{\chi}^-_k\wt{\chi}^0_j\omega}_{\sigma}
g^{\wt{\chi}^0_j\wt{\chi}^-_iW}_{\rho}
C^{\rm INO}_{i}(p_1,p_2,m^2_{\wt{\chi}^-_i},m^2_{\wt{\chi}^0_j},
m^2_{\wt{\chi}^-_k}) \right. \nonumber\\ 
&&\left. \;\;\;\;\;\;\; +
g^{\wt{\chi}^0_i\wt{\chi}^0_kZ}_{\tau}
g^{\wt{\chi}^0_k\wt{\chi}^+_j\omega}_{\sigma}
g^{\wt{\chi}^+_j\wt{\chi}^0_iW}_{\rho}
C^{\rm INO}_{i}(p_1,p_2,m^2_{\wt{\chi}^0_i},m^2_{\wt{\chi}^+_j},
m^2_{\wt{\chi}^0_k}) \right\}\,.
\end{eqnarray}



\end{document}